\documentclass[%
aip,
amsmath,amssymb,
reprint
]{revtex4-1}
\usepackage{graphicx}
\usepackage{dcolumn}
\usepackage{bm}
\usepackage[utf8]{inputenc}
\usepackage[T1]{fontenc}
\usepackage{mathptmx}
\usepackage{etoolbox}
\usepackage{braket}
\usepackage{array,blindtext,enumitem}
\usepackage{color}

\makeatletter
\def\@email#1#2{%
	\endgroup
	\patchcmd{\titleblock@produce}
	{\frontmatter@RRAPformat}
	{\frontmatter@RRAPformat{\produce@RRAP{*#1\href{mailto:#2}{#2}}}\frontmatter@RRAPformat}
	{}{}
}%
\makeatother

\begin{document}
\preprint{APS/123-QED}

\title[Gain-Loss Coupled Systems]{Gain-Loss Coupled Systems}
    \author{Chunlei Zhang}%
	\affiliation{Department of Physics and Astronomy, University of Manitoba, Winnipeg, Manitoba R3T 2N2, Canada}%
	\author{Mun Kim}%
	\affiliation{Department of Physics and Astronomy, University of Manitoba, Winnipeg, Manitoba R3T 2N2, Canada}%
        \author{Yi-Hui Zhang}%
	\affiliation{Department of Physics and Astronomy, University of Manitoba, Winnipeg, Manitoba R3T 2N2, Canada}%
    \author{Yi-Pu Wang}%
	\affiliation{Zhejiang Key Laboratory of Micro-Nano Quantum Chips and Quantum Control, School of Physics, and State Key Laboratory for Extreme Photonics and Instrumentation, Zhejiang University, Hangzhou 310027, China}
    \author{Deepanshu Trivedi}%
	\affiliation{Department of Electrical and Computer Engineering, Florida International University, Miami, Florida 33174, USA}
	\author{Alex Krasnok}%
	\affiliation{Department of Electrical and Computer Engineering, Florida International University, Miami, Florida 33174, USA}
    \author{Jianbo Wang}%
	\affiliation{School of Physical Science and Technology, Lanzhou University, Lanzhou 730000, People’s Republic of China}
    \author{Dustin Isleifson}
	\affiliation{Department of Electrical and Computer Engineering, University of Manitoba, Winnipeg, Manitoba R3T 5V6, Canada}%
    \author{Roy Roshko}
	\affiliation{Department of Physics and Astronomy, University of Manitoba, Winnipeg, Manitoba R3T 2N2, Canada}%
	\author{Can-Ming Hu}
	\affiliation{Department of Physics and Astronomy, University of Manitoba, Winnipeg, Manitoba R3T 2N2, Canada}%
        \email{zhangc23@myumanitoba.ca; hu@physics.umanitoba.ca}
\date{\today}

\begin{abstract}
Achieving oscillations with small dimensions, high power, high coherence, and low phase noise has been a long-standing goal in wave physics, driving innovations across classical electromagnetic theory and quantum physics. Key applications include electronic oscillators, lasers, and spin-torque oscillations. In recent decades, physicists have increasingly focused on harnessing passive oscillatory modes to manipulate these oscillations, leading to the development of diverse gain-loss coupled systems, including photon-photon, exciton-photon, photon-magnon, magnon-phonon, and magnon-magnon couplings. This review provides a comprehensive overview of these systems, exploring their fundamental physical structures, key experimental observations, and theoretical insights. By synthesizing insights from these studies, we propose future research directions to further advance the understanding and application of gain-loss coupled systems for quantum science and quantum technologies. (\textit{The field of gain-loss coupled systems is vast. The authors welcome suggestions and feedback from the community to continuously improve this review article until it is published}).
\end{abstract}

\maketitle 
\section{Introduction}
    Generally, gain denotes the ability to increase a system's energy or amplitude, commonly achieved using devices such as electronic, microwave, and optical amplifiers. It enhances the signal's energy or amplitude without altering properties like frequency, duration, and waveform. In engineering, gain is crucial for communication and radar systems, facilitating long-distance signal transmission with minimal decay and distortion\cite{pozar2011microwave}. It is also employed in feedback systems to counteract decay and maintain stability. In physics, gain is frequently utilized in optics, where an incident beam gains energy through stimulated emission. Gain is fundamental to lasers\cite{murray2018laser}, counteracting losses in the optical cavity to achieve self-oscillation. More broadly, gain is an essential mechanism driving self-oscillations, as modeled by the van der Pol and Rayleigh equations\cite{JENKINS2013167}, often linked to anti-dissipative effects such as negative impedance and negative damping.

    Self-oscillations have been a foundational topic in physics and engineering, continually evolving with new discoveries and technological applications. A well-known example from everyday life is wind instruments, such as trumpets and euphoniums, whose origins can be traced back to ancient bone-carved flutes\cite{conard2009new}. These acoustic self-oscillators are driven by steady air pressure\cite{BrassPhysics}, a principle summarized by Rayleigh in the 19th century\cite{rayleigh1877theory}. Later, van der Pol demonstrated how self-oscillation is realized through negative resistance in electronic experiments\cite{van1927}. Today, the van der Pol oscillator equation serves as a classical model for self-oscillations, widely recognized in physical science, chaotic systems, and biological science, with applications in lasers\cite{murray2018laser}, spin-torque oscillators\cite{slonczewski1996current,berger}, limit cycles\cite{strogatz2015nonlinear}, bifurcations\cite{holmes1978bifurcations}, cardiac cycles\cite{van1928heart}, and neuronal activities\cite{fitzhugh1961impulses}.

    \begin{figure*}[t]
		\centering
		\includegraphics{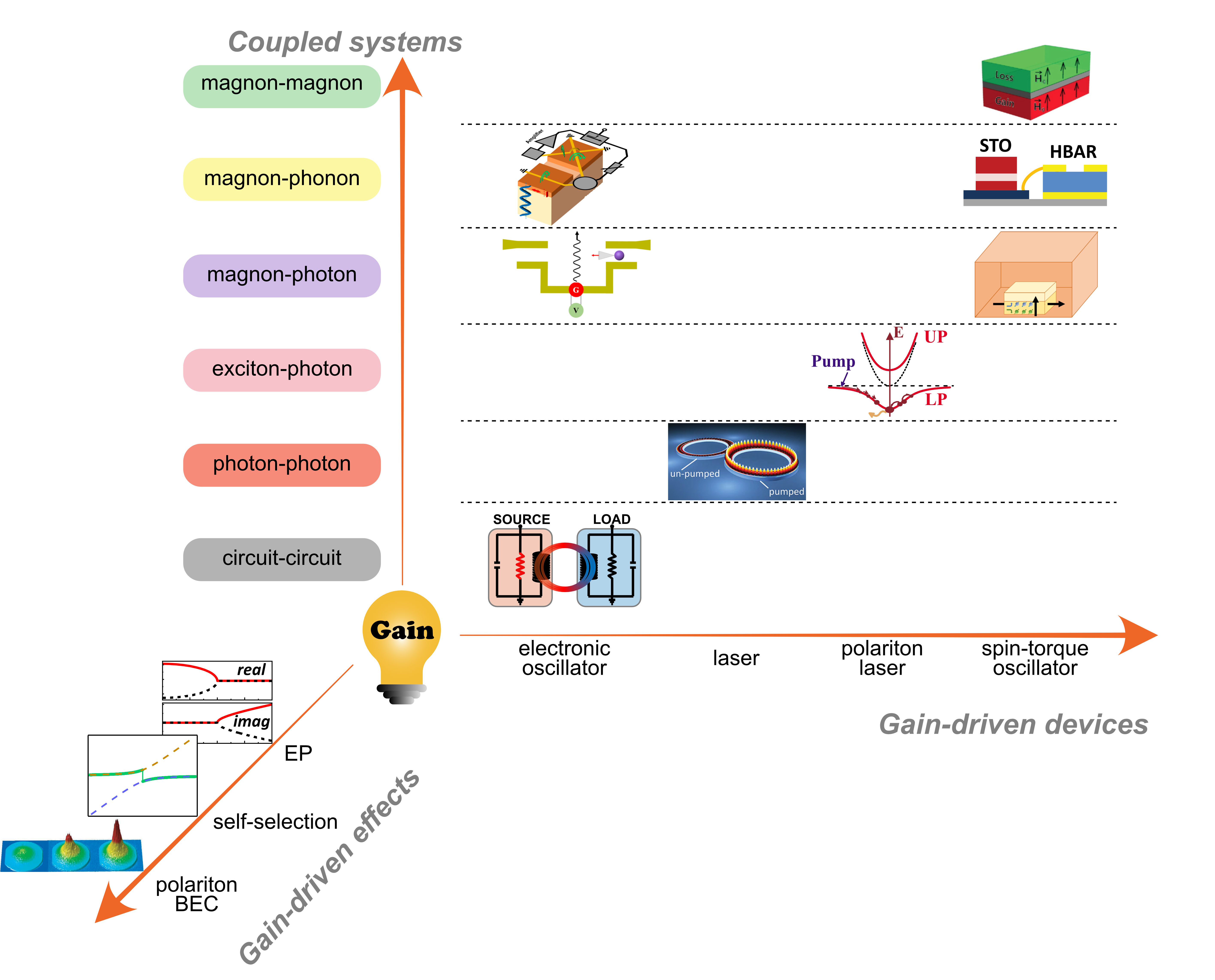}
		\caption{Broad interests in gain-loss coupled systems, including wireless power transfer system, photon-photon\cite{singleTrans}, exciton-photon\cite{kasprzak2006bose}, magnon-photon\cite{GDP,STOmaser}, magnon-phonon\cite{torunbalci2018magneto}, and magnon-magnon coupling\cite{PTmagnon}. It advances the interdisciplinary development of various gain-driven devices such as electronic oscillators, polariton lasers, conventional lasers, and spin-torque oscillators. The exploration of these systems facilitates the understanding and utilization of fascinating effects like polariton BEC\cite{kasprzak2006bose}, EP\cite{wang2020steering}, and self-selection\cite{GDP}. Photon-photon coupling inset is reproduced with permission from Hodaei et al., Laser Photonics Rev. 10, 494–499(2016). Copyright 2016, John Wiley and Sons. Exciton-photon coupling inset is adapted with permission from Ref.~\onlinecite{DengPolaritonBEC}. Copyrighted by the American Physical Society. Magnon-photon coupling and self-selection insets are adapted with permission from Ref.~\onlinecite{GDP} and~\onlinecite{STOmaser}. Copyrighted by the American Physical Society. Right inset of magnon-phonon coupling is reproduced with permission from Torunbalci et al., Sci. Rep. 8, 1119 (2018). Copyright 2018, licensed under a Creative Commons Attribution (CC BY) License. Left inset of magnon-phonon coupling is reproduced with adapted from Ref.~\onlinecite{maoscillator}. Copyrighted by the American Physical Society. Magnon-magnon coupling inset is reproduced with permission from Ref.~\onlinecite{PTmagnon}. Copyrighted by the American Physical Society. Polariton BEC inset is reproduced with permission from Kasprzak et al., Nature 443, 409–414 (2006). Copyright 2006, Springer Nature. EP inset is reproduced with permission from Wang et al., Nat. Commun. 11, 5663 (2020). Copyright 2020, licensed under a Creative Commons Attribution (CC BY) License.}
		\label{fig1}
    \end{figure*}

    The interaction between self-oscillation and lossy oscillation results in a gain-loss coupled system. Due to the gain, these coupled systems can sustain steady hybridized oscillations. By tuning the gain, loss, coupling strength, and frequency detuning between gain and lossy modes, hybridized oscillations can be precisely manipulated. This leads to novel phenomena not observed in isolated gain-driven oscillators, such as exceptional points (EPs)\cite{el2018non,ozdemir2019parity,Ashida}, polariton Bose-Einstein condensates (BECs)\cite{DengPolaritonBEC,byrnes2014exciton}, and self-selection of gain-driven polaritons\cite{GDP}.

    An EP denotes the non-Hermitian degeneracy in a gain-loss coupled system. Unlike conservative coupled systems with real-valued frequencies, gain and loss in these systems are associated with imaginary coefficients, rendering the Hamiltonian non-Hermitian and resulting in complex frequencies\cite{el2018non,ozdemir2019parity,Ashida,Federico2022}. At an EP, gain and loss compete with coupling, leading to a degenerate real eigenfrequency\cite{Heiss2004physics,berry2004physics,Heiss_2012}, in contrast to the frequency repulsion in conservative systems. Coupled systems featuring EPs have demonstrated capabilities in nonreciprocal transmission\cite{PTphononmedia,doppler2016dynamically, wang2020steering}, loss-induced lasing\cite{revival}, coherent perfect absorption\cite{absorber, twomodelase, wong2016lasing, fleury2015invisible}, and mode suppression\cite{feng2014single,single,optoe,liu2018observation}. These properties have inspired advancements in various fields, including wireless power transfer\cite{assawaworrarit2017robust, song2021wireless}, non-Hermitian optics\cite{ozdemir2019parity,Wang23}, non-Hermitian acoustics\cite{PTacoustic, huang2024acoustic}, and non-Hermitian magnonics\cite{Hurst2022,YUAN20221,YU20241}.
    
    Exciton-polariton BECs hold significant potential as innovative photonic sources\cite{ZHANG2022100399}. Exciton polaritons\cite{SAVONA1995733,PolaritonDicke} are coupled states of cavity photons and excitons\cite{nonlinearExciton} (bound pairs of electrons and holes). The gain and loss in these systems are caused by external pumping and cavity dissipation\cite{SAVONA1995733}, respectively. Under high-power pumping, condensed polaritons in this non-conservative system are recognized as non-equilibrium BECs due to their macroscopic ground-state occupation\cite{kasprzak2006bose,balili2007bose,DengPolaritonBEC,byrnes2014exciton}. Their photon emission shows high coherence and output power, making them novel optical sources known as polariton lasers\cite{polaritonlaser,solidPolaLaser}.

    Beyond exciton-polaritons, gain-driven polaritons exhibit intriguing phenomena, particularly the self-selection of a single bright mode. This phenomenon occurs when collective spin excitations (magnons) interact with microwave cavity photons, leading to the spontaneous selection of one dominant oscillatory mode from two coupled eigenmodes\cite{GDP,kim2024low}. This self-selection results in systems that demonstrate high power, high coherence, and sharp emission linewidth. Furthermore, advancements in magnon-phonon systems, utilizing on-chip high-frequency acoustic resonators, have achieved low-noise phonon emission\cite{maoscillator}. Despite ongoing research and evolving understanding, these classical hybrid systems hold significant potential for applications in on-chip coherent microwave sources and amplifiers.

    As summarised in Fig. \ref{fig1}, the field of gain-loss coupled systems encompasses a wide range of phenomena and applications. Numerous systems are involved, inspiring the development of various gain-driven devices. In photon-photon and magnon-magnon systems, scientists focus on EPs of systems with balanced gain and loss, known as parity-time (PT) symmetric systems. These systems have Hamiltonians invariant under PT transformation. In hybrid systems, such as exciton-photon, magnon-photon, and magnon-phonon systems, phenomena like polariton BECs and self-selection are observed. Compared with the PT-symmetry, these effects emerge from the nonlinearity caused by the gain-induced high-amplitude oscillation. 

    Given the expansive interpretation of gain in non-Hermitian physics, this article is grounded in explicit gain, wherein the amplitude of a resonant mode is modulated by an exponential function. It reviews gain-loss coupled systems through both PT-symmetry and nonlinearity, elucidating their connections and distinctions. First, in Sec. \ref{sec2}, we introduce the concept of gain-driven oscillation by examining typical self-oscillations, including electronic oscillators, lasers, and spin-torque oscillations. Then, in Sec. \ref{sec3} we focus on gain-loss coupled systems, identifying theoretical models and the corresponding gain-driven effects, and highlighting their potential for applications. Finally, in Sec. \ref{sec4}, we discuss the future development and potential advances of gain-loss coupled systems.

\section{Gain-driven harmonic oscillation} \label{sec2}
    \begin{table*}
	\caption{\label{tab:table1}Overview of basic elements, schematics, and features of various gain-driven devices. In electronic oscillators, gain is achieved through negative resistance provided by an amplifier. Semiconductor lasers utilize gain from stimulated emission in a gain medium. Spin-torque oscillators achieve gain via spins, with the resulting spin torque counteracting damping torque.}
    \centering
	\begin{ruledtabular}
    \centering
		\begin{tabular}{c m{0.3\textwidth}<{\centering} m{0.2\textwidth}<{\centering} m{0.25\textwidth}<{\centering}}
			Devices&Basic elements&Schematics&Features\\ \hline
   
			Electronic oscillator  & \begin{tabular}{m{0.3\textwidth}<{\centering}}\begin{itemize}[leftmargin=1em,noitemsep,topsep=1em,parsep=4pt,partopsep=0pt] \item Circuit resonator \item Amplifier-based negative resistance \item Voltage cap of amplifier \end{itemize}\end{tabular} & \begin{tabular}{m{0.2\textwidth}}\begin{center}\includegraphics[scale=1]{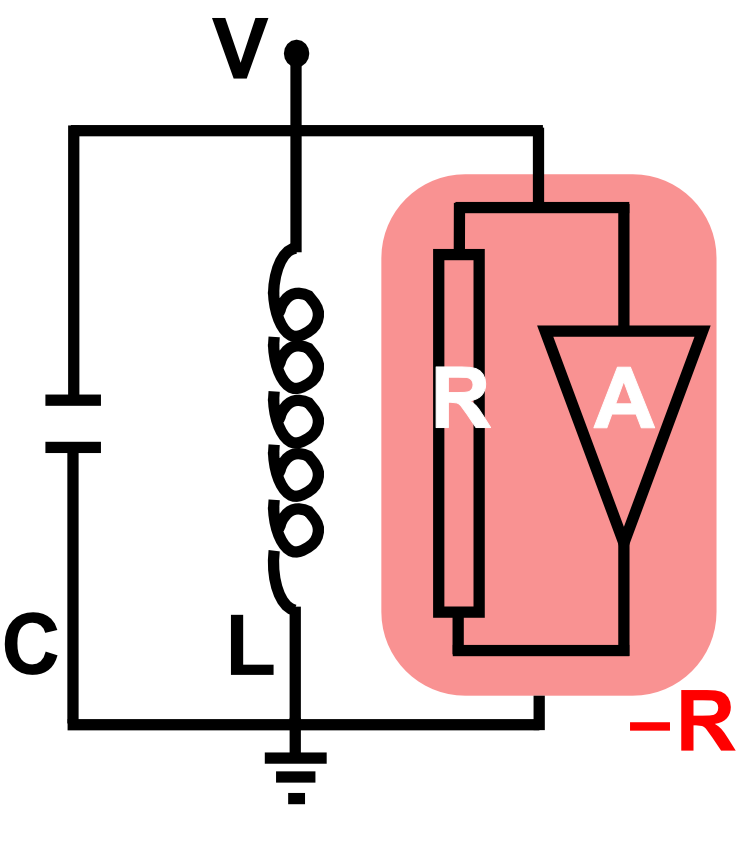}\end{center}\end{tabular} & \begin{tabular}{m{0.25\textwidth}<{\centering}}\begin{itemize}[leftmargin=1em,noitemsep,topsep=1em,parsep=4pt,partopsep=0pt] \item Centimeter scale \item Large frequency range \end{itemize}\end{tabular} \\ \hline
			
            Semiconductor laser & \begin{tabular}{m{0.3\textwidth}<{\centering}}\begin{itemize}[leftmargin=1em,noitemsep,topsep=1.2em,parsep=4pt,partopsep=0pt] \item Optical cavity \item Gain medium \item Gain saturation \end{itemize}\end{tabular} & \begin{tabular}{m{0.2\textwidth}}\begin{center}\includegraphics[scale=1]{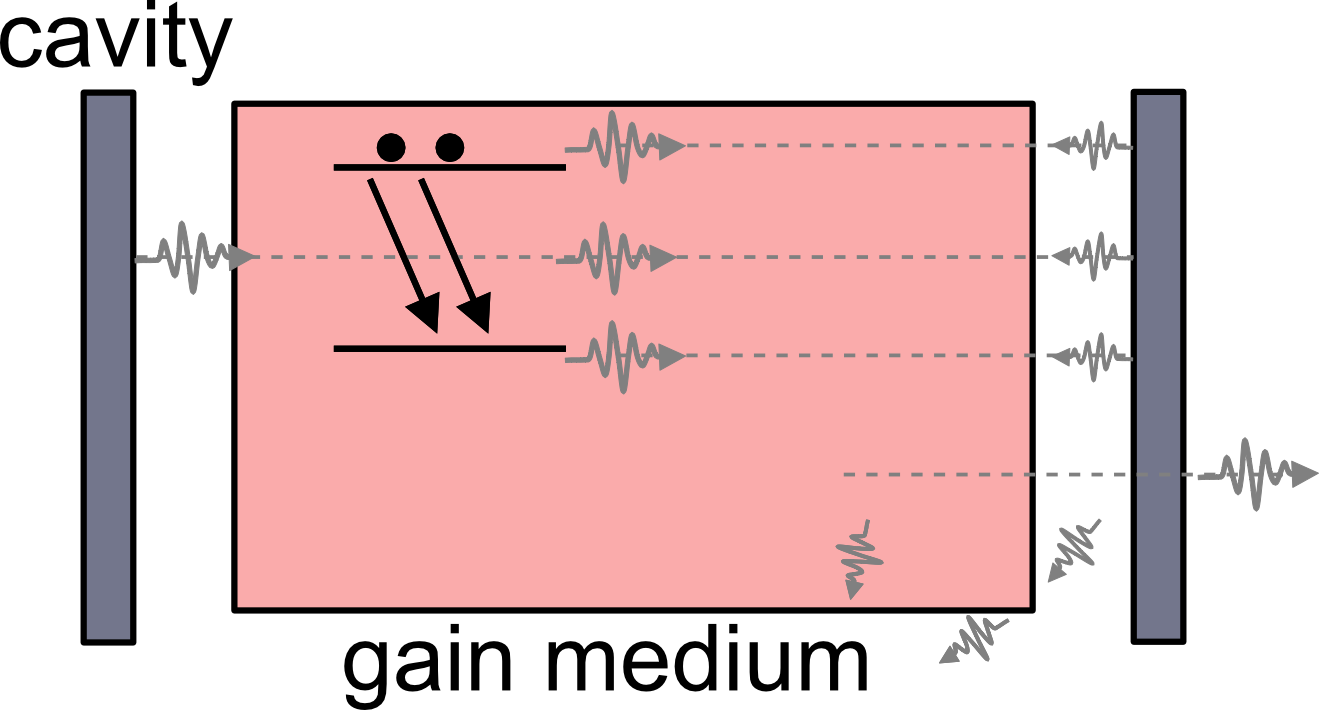}\end{center}\end{tabular} & \begin{tabular}{m{0.25\textwidth}<{\centering}}\begin{itemize}[leftmargin=1em,noitemsep,topsep=1.2em,parsep=4pt,partopsep=0pt] \item Millimeter scale \item Coherent emission \item Directional propagation \end{itemize}\end{tabular} \\ \hline
			
            Spin-torque oscillator\cite{covington2005ringing} & \begin{tabular}{m{0.3\textwidth}<{\centering}}\begin{itemize}[leftmargin=1em,noitemsep,topsep=1.2em,parsep=4pt,partopsep=0pt] \item Larmor precession \item Spin torque \item Large precession angle \end{itemize}\end{tabular} & \begin{tabular}{m{0.2\textwidth}}\begin{center}\includegraphics[scale=1]{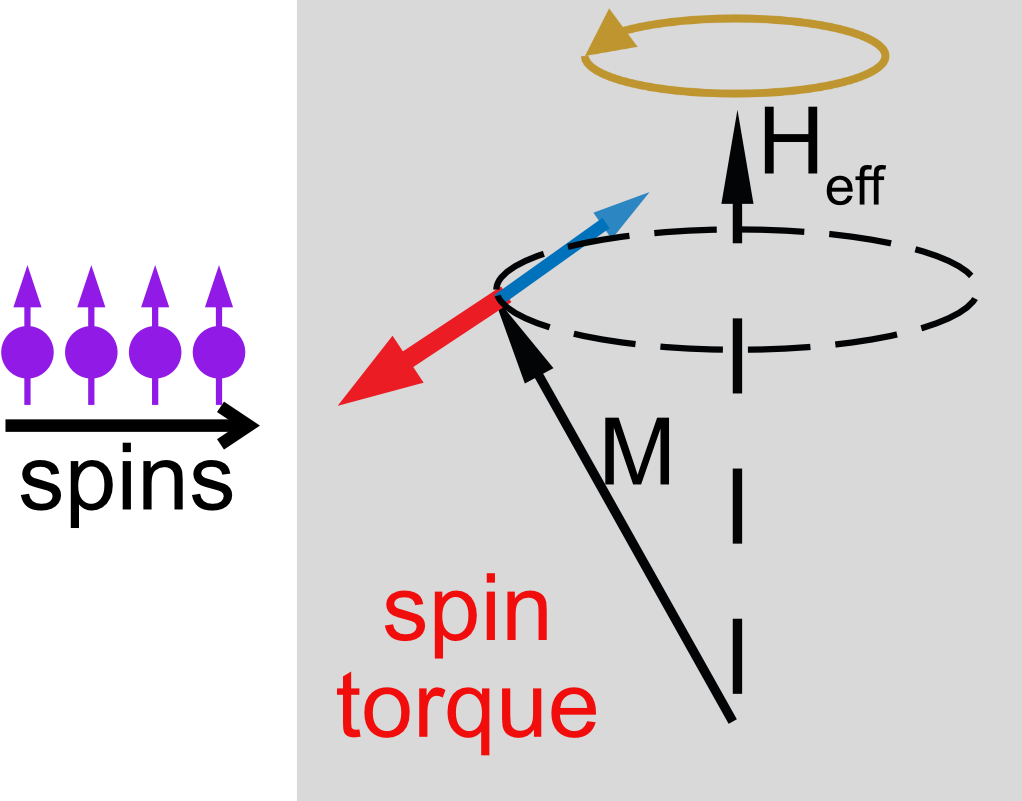}\end{center}\end{tabular} & \begin{tabular}{m{0.25\textwidth}<{\centering}}\begin{itemize}[leftmargin=1em,noitemsep,topsep=1.2em,parsep=4pt,partopsep=0pt] \item Nanometer scale \item Adjustable frequency \end{itemize}\end{tabular} \\
		\end{tabular}
	\end{ruledtabular}
\end{table*}
    
    We define harmonic oscillations driven by gain as gain-driven harmonic oscillations, which include self-oscillations and parametric oscillations. Unlike the classical forced oscillation model, where the driving force matches the system's natural frequency, self-oscillators and parametric oscillators do not need external motivations aligning with the system's resonance frequency. This interesting intriguing aspect was first recognized by Rayleigh. In both types of oscillators, gain is essential to maintain oscillation, thereby classifying them as gain-driven oscillations. Our discussion focuses on gain-driven oscillations of self-oscillators in Sec. \ref{sec2A}, while the gain-driven oscillation through parametric pumping is briefly introduced in Sec. \ref{sec2B}. We also introduce the novel concept of virtual gain and its recent developments in Sec. \ref{sec2C}.

    \subsection{Gain-driven oscillations of self-oscillators} \label{sec2A}
    
    In this section, we will present gain-driven oscillations of electronic, optical, and magnetic self-oscillators, summarized in Table \ref{tab:table1}. Generally, these oscillations require mechanisms that convert power into negative damping of an oscillator, resulting in an amplified oscillation amplitude. The rate at which the amplitude increases over time is characterized as gain, a special form of dissipation opposite to the common dissipation known as loss. However, oscillatory systems with gain cannot be simplistically modeled as harmonic oscillators with unlimited energy increases. The amplitude of a realistic gain-driven oscillation always gradually escalates to be steady.
    
    Mathematically, the dynamics of a general gain-driven oscillator $a(t)$ can be typically modeled using a first-order van der Pol equation\cite{yamamoto1983and,strogatz2015nonlinear,slavin2009nonlinear,Zhang},
    \begin{equation} \label{Eq1}
		\frac{da}{dt}=-i\omega_{0}a+(G-\gamma|a|^{2})a,
    \end{equation}
    where three basic elements are necessary to realize gain-driven oscillation\cite{JENKINS2013167}. First, the system should incorporate a resonator, indicated by the angular frequency $\omega_{0}$. Second, the system should incorporate the negative damping to activate the oscillation, indicated by the gain coefficient $G$ ($G>0$). Third, the van der Pol nonlinearity, indicated by $\gamma|a|^{2}$, should be incorporated to stabilize the oscillation (or clamp the oscillation amplitude) by counteracting the gain coefficient.
    
    Gain-driven oscillators can lock onto injected resonant signals\cite{odyniec2002rf,RazaviIJL,generalIJL}, a phenomenon known as injection locking. This allows the oscillator to synchronize with a range of signal frequencies while maintaining a nearly constant output. This locking behavior is distinctly different from the forced oscillation of damped oscillators, whose amplitude can only reach a maximum at the resonant frequency. Besides, it should be noted that in practical cases, the oscillation frequencies of the gain-driven systems are influenced by amplitude, leading to Duffing\cite{strogatz2015nonlinear,landau1960mechanics} or Kerr nonlinearity\cite{kerr1875xl, agrawal2000nonlinear,slavin2009nonlinear,nonlinearmagnon}. However, in this section, we focus solely on gain-driven oscillations with van der Pol nonlinearity. The following paragraphs provide a detailed analysis of three typical gain-driven oscillators, illustrating how negative damping is realized in each.
                                      
    In electronics, negative damping is achieved through negative resistance\cite{pozar2011microwave}. In contrast to common resistance that lowers voltage, negative resistance should increase it, which can be realized by an electronic amplifier. By embedding the amplifier into the circuit resonator, the voltage oscillation amplitude rises over time until it stabilizes at the amplifier's maximum voltage. Thanks to advanced electronic manufacturing techniques, we can engineer electronic oscillators that operate across a broad radio frequency range\cite{ishak1988magnetostatic,odyniec2002rf}, from Hz to dozens of GHz, while keeping the devices compact, typically at a centimetre scale. This versatility makes them absolutely indispensable in the electronic products of modern society.

    In semiconductor lasers, negative damping is achieved through the gain medium. Semiconductor lasers, known as laser diodes, are typically composed of a millimetre-sized semiconductor optical cavity and its inside medium\cite{roomlaser,haken1984laser}. When the medium is charged to the state of population inversion, incident cavity photons will stimulate the emission of additional photons, resulting in coherent light amplification. This process reaches stability at gain saturation\cite{yamada2014theory, murray2018laser}, where the net gain becomes zero, leading to a steady-standing oscillation within the cavity. This coherent standing wave partially radiates outside the cavity, resulting in the directional coherent light emission. It shows extensive applications across various domains, such as laser cutting\cite{bromberg1995laser} in industry, laser surgery\cite{surgery} in medication, optical disk\cite{disk} in entertainment, and laser cooling\cite{HANSCH197568,cool1979} in scientific research.

    In spin-torque oscillators, the negative damping is generated through the interaction between spins and magnetization. Spin-torque oscillators are nanoscale semi-classical magnetic oscillators utilizing microscopic spins to drive the oscillation of macroscopic magnetization\cite{slonczewski1996current, berger,sotorque}. Typically, the Larmor precession of macroscopic magnetization is dissipated by a damping torque. However, microscopic spins, generated by current passing through magnetic or heavy metal nanoscale thin films, create a torque opposing the damping torque of the magnetization vector. Consequently, spins cause a negative damping activating the magnetic precession\cite{kiselev2003microwave,sotorque,deac2008bias,Harms2022}. This dynamics can be stabilized at a large precession angle, recognized as the nonlinear phenomena of magnetic resonance\cite{slavin2009nonlinear,STNOieee}. Since the Larmor precession can be manipulated through a magnetic field, this magnetic gain-driven oscillator can achieve an adjustable output frequency\cite{LinALP2010}. Spin-torque oscillators show potential as a new platform for the microwave source\cite{kaka2005mutual,deac2008bias,sotorque}, and computing\cite{torrejon2017neuromorphic,isingSTO}.

    Presently, all three typical gain-driven oscillations continue to attract the interest of physicists and engineers. While the common feature, injection locking, is widely observed in these systems\cite{Paciorek,ijlSTO,laserIJL}, each has its own developmental direction and unique challenges. For instance, spin-torque oscillators face issues such as low-power emission and broad emission linewidth\cite{FluctuationsSTO}, challenges not typically encountered by electronic oscillators and semiconductor lasers. Thus, comparing and understanding gain-driven harmonic oscillations in different systems offers unique insights and opportunities for innovation.
    
    \subsection{Gain-driven oscillations through parametric pump} \label{sec2B}
    Oscillations realized through high-frequency parametric pumping are another type of gain-driven oscillation\cite{gainPA}. In this scenario, the gain is obtained through the external driving whose frequency is significantly higher than the system's natural resonant frequency. In classical mechanics, this phenomenon is known as parametric resonance\cite{landau1960mechanics}, often illustrated by a swing driven by a periodic vertical force at twice its natural frequency\cite{swing}. Generally, these dynamics involve nonlinear processes that down-convert high-frequency oscillations to resonate with the system’s natural frequency, thus they are not restricted to double-frequency excitation\cite{elePA,QASAER,optoMech}.
    
    In electronics and optics, this principle is utilized to achieve parametric amplifiers, where an electronic (optical) injected signal is amplified through a nonlinear circuit (nonlinear medium) pumped by a high-frequency signal\cite{Weiss,elePA,Meystre2007,hao2020optoelectronic}. Given this amplification effect, combining an optical parametric amplifier within an optical cavity results in an optical parametric oscillator\cite{OpticalParametric}. Since this type of oscillation is also achieved through the gain, it is classified as a gain-driven oscillation.

    While parametric pumping is similar to the electronic amplifier, a critical distinction is that this gain depends on the phase relationship between the pumping frequency and the resonant frequency\cite{grimshaw1990nonlinear}. This phase dependency makes parametric gain-driven oscillations distinct from those described by a van der Pol equation.
    
    \subsection{Gain-driven oscillations through virtual gain} \label{sec2C}
    Compared to the traditional gain mechanisms of self-oscillators and parametric pumping, virtual gain can be applied to a resonant mode without requiring intrinsic amplification. As discussed in Sec. \ref{sec2A} and \ref{sec2B}, traditional gain arises from specially designed resonators with inherent amplification. For instance, in self-oscillators like lasers, the gain comes from the gain medium within the cavity, whereas in parametric pumping, it originates from frequency conversion based on the system’s intrinsic nonlinearity. In contrast, virtual gain-driven oscillation results from an external periodic excitation signal whose amplitude decays exponentially over time, bypassing the need for intrinsic amplification.

    Virtual gain is a novel and intriguing concept. While initially introduced in optics and often explained using complex terminology, its core principle can be first understood through the lens of classical forced damped oscillations, as covered in standard textbooks. Consider a resonant mode driven by a force with a complex frequency. The forced damped oscillation is described by the equation:
    \begin{equation}
        \frac{d{a}}{dt}=-(i{{\omega }_{0}}+{{\kappa }}){{a}}+{{s}_{e}}{{e}^{-i({{\omega }_{r}}-iG)t}},
        \label{Eqvg}
    \end{equation}
    where $\omega_{0}$ represents the angular frequency of the resonator mode, and $\kappa$ denotes the loss coefficient. The driving force has an angular frequency ${\omega }_{r}$ and decays at a rate $G$ ($G > 0$), while $s_{e}=\sqrt{2\kappa_{E}}s_{0}$ is a constant related to the force's amplitude and external dissipation. Given an initial amplitude of resonator $a(0)=s_{e}$, three outcomes can be observed: (1) When no driving force is applied, $s_{e}=0$, the system exhibits simple damped oscillation with a decay rate of $\kappa$; (2) When the driving force decays slowly, $G \leq \kappa$, the resonator absorbs energy from the force, indicating the amplitude of the resonator remains no larger than that of the driving force; (3) When the driving force decays rapidly, $G > \kappa$, this force decays faster than the resonator mode, effectively giving the resonator a ``virtual'' gain characterized by $G_{v} = G - \kappa$, where the resonator behaves as though it were actively amplifying the oscillation of the force. A detailed derivation of virtual gain is provided in Appendix \ref{APPvirtualG}.

    In optics, virtual gain occurs when the amplitude of the scattered signal exceeds that of the incident field\cite{VPTS2020,zouros2024anisotropic,FXGuan2023,guan2024compensating,gu2022transient,Seunghwi2023}, while the incident field, corresponding to the external force in Eq. \ref{Eqvg}, is named as complex frequency excitation. This condition is achievable when the resonator mode is initially populated and the excitation decays faster than the mode itself. Under these circumstances, the ratio of the scattered to incident signal can surpass one and may even diverge as the system approaches the resonator's eigenmode. By utilizing virtual gain, researchers have unlocked numerous phenomena in the ``virtual'' regime, such as virtual perfect absorption\cite{Krasnok2019,Baranov2017}, virtual critical coupling \cite{VCC2020,hinney2024efficient}, virtual PT symmetry\cite{VPTS2020}, and optical pulling forces\cite{Lepeshov20}. Hinney et al. recently demonstrated efficient light transfer in integrated photonic devices by precisely tailoring the excitation signal over time\cite{hinney2024efficient}. In elastodynamics, this technique has enabled efficient absorption and transmission of elastic waves through coherent virtual absorption\cite{2019Trainiti}, building on the principle of virtual absorption. Similarly, virtual gain has been applied in superlensing and metamaterials for sub-diffraction imaging, where complex decaying signals compensate for losses, enhancing imaging resolution beyond diffraction limits\cite{FXGuan2023}. In molecular sensing, complex-frequency waves have amplified detection of molecular vibrations, recovering vibrational modes that would otherwise be lost to damping\cite{zeng2024synthesized}. A recent report shows that virtual gain involves transforming passive anisotropic media into amplifying systems, allowing active control of particle scattering by adjusting incident radiation, without altering the medium's intrinsic gain\cite{zouros2024anisotropic}.

\section{Gain-loss coupled systems}\label{sec3}
    Based on these gain-driven oscillations, researchers can realize gain-loss coupling in various systems, showing featured effects. A general gain-loss coupled system incorporates two interacting oscillatory modes: the gain-driven mode and the damped mode. Both modes are influenced by the damping and negative damping of the environment, represented by loss and gain, respectively. Two aspects are of particular concern: first, the competition between the environment and coupling\cite{el2018non,ozdemir2019parity,Federico2022}, resulting in frequency degeneracy at the EP; second, the high-amplitude oscillation induces nonlinearity, potentially resulting in phenomena like polariton BECs\cite{kasprzak2006bose} and self-selection\cite{GDP}. Currently, studies on gain-loss coupled systems can be classified into two models depending on the focus: linear gain-loss coupled model, which emphasizes physics around EPs, and nonlinear gain-loss coupled model, which is often used in describing steady-state polaritons. Our discussion in Sec. \ref{sec3A} briefly introduces the theories of these two primary models of gain-loss coupled systems. We then review the corresponding systems and their applications in Sec. \ref{sec3B} and Sec. \ref{sec3C}, respectively. Notably, virtual gain serves as a novel technique to realize gain-driven oscillations without relying on intrinsic amplification processes; thus, Section \ref{sec3D} is dedicated to introducing the emerging field of virtual gain-loss coupled systems.
    
\subsection{Theories}\label{sec3A}
    To understand the physics of gain-loss coupled systems, selecting the appropriate theoretical model provides significant convenience. Unlike conservative systems, where both modes experience negligible dissipation, gain-loss coupled systems must account for two additional factors: the competition between environmental dissipation and coupling strength, and the nonlinearity induced by amplified oscillation. These considerations respectively characterize the linear and nonlinear gain-loss Hamiltonians.

    \begin{figure}[tbp]
		\centering
		\includegraphics{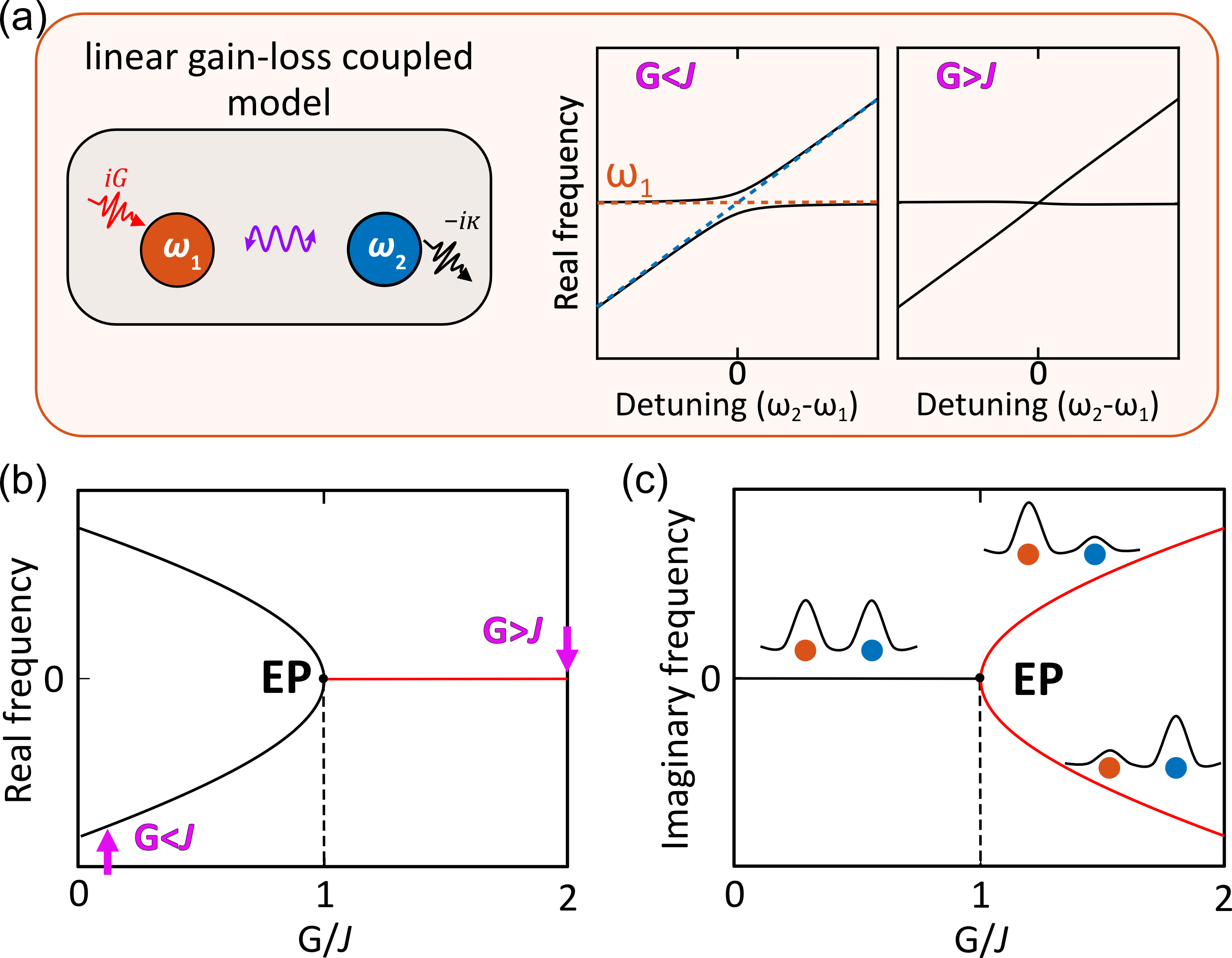}
		\caption{(a) Diagram of a linear balanced gain-loss coupled model, showing avoided crossing and frequency degeneracy at dispersion relations. Schematically, (b) real and (c) imaginary parts of the complex frequency as functions of the gain to coupling strength ratio $G/J$, highlighting the EP. The three insets in the right panel denote the hybridization ratios of different coupled eigenmodes.}
		\label{fig2}
    \end{figure}

    For the linear gain-loss coupled systems, the focus is on how the degeneracy of the coupled system is significantly influenced by environmental dissipation, characterized as gain and loss. This topic is currently a focal point in the study of coupling physics, including optics, magnonics, and acoustics. In a simple case, a linear gain-loss coupled system can be represented by two zero-detuning complex frequency modes, $\omega_{0}+iG$ and $\omega_{0}-i\kappa$ ($G,\kappa>0$ corresponds gain and loss), coupled through a real-valued strength $J$. Typically, the values of gain and loss are hard to be identical, but a gauge transformation shifts the reference value of the imaginary part from zero to a reference defined as $\chi=(G-\kappa)/2$. With this reference, the dynamic matrix of this coupled system can be written as follows,
    \begin{equation}
		\begin{aligned}
			H=&\left(
			\begin{aligned}
				\omega_{0}+i\beta \ \ \ \ \ \ J \ \ \ \ \\
				 \ \ \ \ J \ \ \ \ \ \ \ \omega_{0}-i\beta
			\end{aligned}\right).
		\end{aligned}
		\label{Eq2}
    \end{equation}
    where the coefficient $\beta=(G+\kappa)/2$ refers to the balanced gain and loss with respect to reference $\chi$. We refer to this phenomenological dynamic matrix as the general form of a linear gain-loss Hamiltonian, capable of describing a wide range of gain-loss coupled systems, from classical to quantum regimes. It remains consistent with quantum systems by setting $\hbar=1$.
    
    The dynamic matrix of this linear gain-loss coupled system inherently exhibits PT symmetry, with the parity transformation involving an exchange of the diagonal terms and the time transformation involving a swap of loss with gain. Therefore, linear gain-loss coupled systems are broadly accepted as PT-symmetric systems.
    
    To clarify terminology, the term ``PT-symmetric'' applies broadly to non-Hermitian systems that exhibit complex eigenvalues, extending beyond those comprising resonant modes with gain and loss. Originally proposed by Bender et al.\cite{bender}, PT-symmetry allows complex eigenvalues in a non-Hermitian Hamiltonian without explicit dissipation. Additionally, as introduced in Ref.~\onlinecite{ozdemir2019parity}, a passive non-Hermitian system can be transformed to possess two complex eigenvalues with positive and negative imaginary parts, representing gain and loss modes, respectively. However, in this article, we focus on specific cases where the gain explicit gain results from an exponentially rising amplitude, as described in Sec. \ref{sec2}. Hence, to distinguish these systems from the broader class of PT-symmetric systems, we introduce the term ``linear gain-loss coupled systems''. This term is specifically intended to refer to cases involving PT-symmetric lasers and magnonics with explicit gain.

    In linear gain-loss coupled systems, environmental dissipation induces a transition of eigenvalues. We set $G=\kappa$ to exhibit a concise example as shown in Fig. \ref{fig2}(a). When the environmental dissipation is relatively small and featured as strong coupling ($G<J$), the system exhibits an avoided crossing pattern, similar to what is commonly seen in conservative systems. However, when the environmental dissipation is large and featured as weak coupling ($G>J$), the system shows a crossing pattern with a degenerated oscillation frequency at zero detuning.

    To characterize the competition between coupling strength and gain, the phase transition of eigenvalues from strong to weak coupling is manifested as a function of $G/J$ and the eigenvalues. As shown in Fig. \ref{fig2}(b) and (c), this transition is illustrated through the features of the complex eigenfrequency, leading to the PT-symmetric and PT-broken phases. In the real part of the eigenfrequency, the PT-symmetric phase shows two distinct coupled oscillating frequencies, while the PT-broken phase shows a single coupled oscillating frequency. In the imaginary part of the eigenfrequency, the PT-symmetric phase shows two coupled eigenmodes with zero dissipation, whereas in the PT-broken phase, one mode is amplified and the other decays. Within the PT-broken phase, the coupling is so weak that the two modes can't equally hybridize. The transition point, defined as $G=J$, is known as the EP and is characterized by the degeneracy of eigenfrequencies. Attracted by these intriguing features around the EP, many gain-loss coupled systems are designed to be PT-symmetric\cite{single,assawaworrarit2017robust,wang2020steering,el2018non,sciencePToptics,ozdemir2019parity,Hurst2022}.
    \begin{figure}[tbp]
		\centering
		\includegraphics{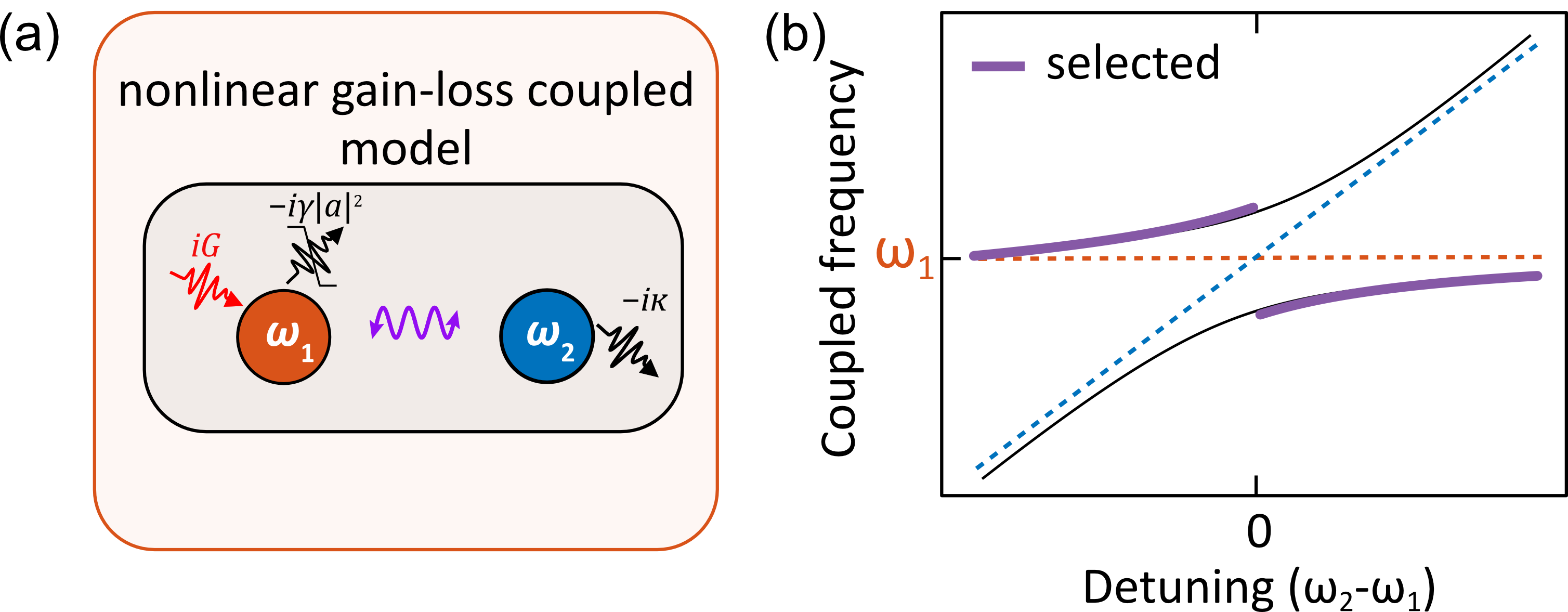}
		\caption{(a) Diagram of a nonlinear gain-loss coupled model. (b) Corresponding schematic frequency dispersion indicated by purple curves, featuring the existence of a single coupled mode.}
		\label{fig3}
    \end{figure}

    In the gain-loss coupled system, the gain usually induces a high-amplitude oscillation, making the system being nonlinear. We refer to the systems that include both environmental dissipation, coupling, and nonlinearity as the nonlinear gain-loss coupled systems. Following the discussion in Sec. \ref{sec2A}, a concise example can be conducted by including both environmental gain, loss, coupling strength, and van der Pol nonlinearity. As shown in Fig. \ref{fig3}(a), the general form of the Hamiltonian for such a coupled system is given by
    \begin{equation}
		\begin{aligned}
			H=&\left(
			\begin{aligned}
				\omega_{1}+iG-i\gamma|a|^{2} \ \ \ \ \ \ \ \ J \ \ \\
				J \ \ \ \ \ \ \ \ \ \ \omega_{2}-i\kappa
			\end{aligned}\right),
		\end{aligned}
		\label{Eq3}
    \end{equation}
    where $\omega_{1,2}$ are the uncoupled frequency, $\gamma$ is the van der Pol coefficient caused by the electronic oscillator, and the gain coefficient should be $G>\kappa$ to ensure the activation of the system. In this system, nonlinearity induces a self-selection effect on the eigenmodes. As shown in Fig. \ref{fig3}(b), using the analysis method proposed by Yao et al.\cite{GDP}, the coupled frequency of the nonlinear gain-loss coupled system denoted by the purple color is compared with that of the strong coupled linear gain-loss system. It indicates that the nonlinearity induced by the high-amplitude oscillation will select the mode close to the frequency of the gain-driven oscillator as the sole final state, while the other mode is dissipated. It can be observed that only the oscillatory branch close to $\omega_{1}$ is sustained.

    The gain-loss hybrid systems can be described by nonlinear gain-loss coupled models. The most renowned gain-loss hybrid system is the exciton-polariton system, commonly known as the driven-dissipative system. This system is driven through the gain of parametric pumping, while its energy decays in the form of light luminescence\cite{kasprzak2006bose,ddphysics,Wouters}. This system is composed of particles of interacting excitons and photons. Under a high power pumping, the excited particles inside the system have a large population exceeding the density threshold, resulting in a condensation to the ground energy. The condensate caused by a large population is dominated by the Kerr-type nonlinearity\cite{Wouters}, where details will be introduced in the following section.
    
    Linear and nonlinear models are different methodologies for studying gain-loss coupled systems. Importantly, these models are not contradictory. Since the competition between coupling strength and environmental dissipation always remains, the nonlinear gain-loss coupled system can also exhibit PT-symmetry and EPs\cite{assawaworrarit2017robust,SAVONA1995733,gao2015observation,Deng2023,weis2022exceptional}. A detailed derivation of the dynamics of the gain-loss coupled system is provided in Appendix \ref{AppSECH}.
    
\subsection{Linear gain-loss coupled systems}\label{sec3B}
    Focusing on properties around EPs, gain-loss coupled systems are characterized by the competition between gain, loss, and coupling strength. EPs can be realized without nonlinearity, making these systems substantial platforms for exploring PT-symmetry. In the following discussion, we will review the development, innovations, and challenges of PT-symmetric systems across the fields of lasers and magnonics.
    
    \subsubsection{PT-symmetric laser systems}
    \begin{figure}[htbp]
		\centering
		\includegraphics{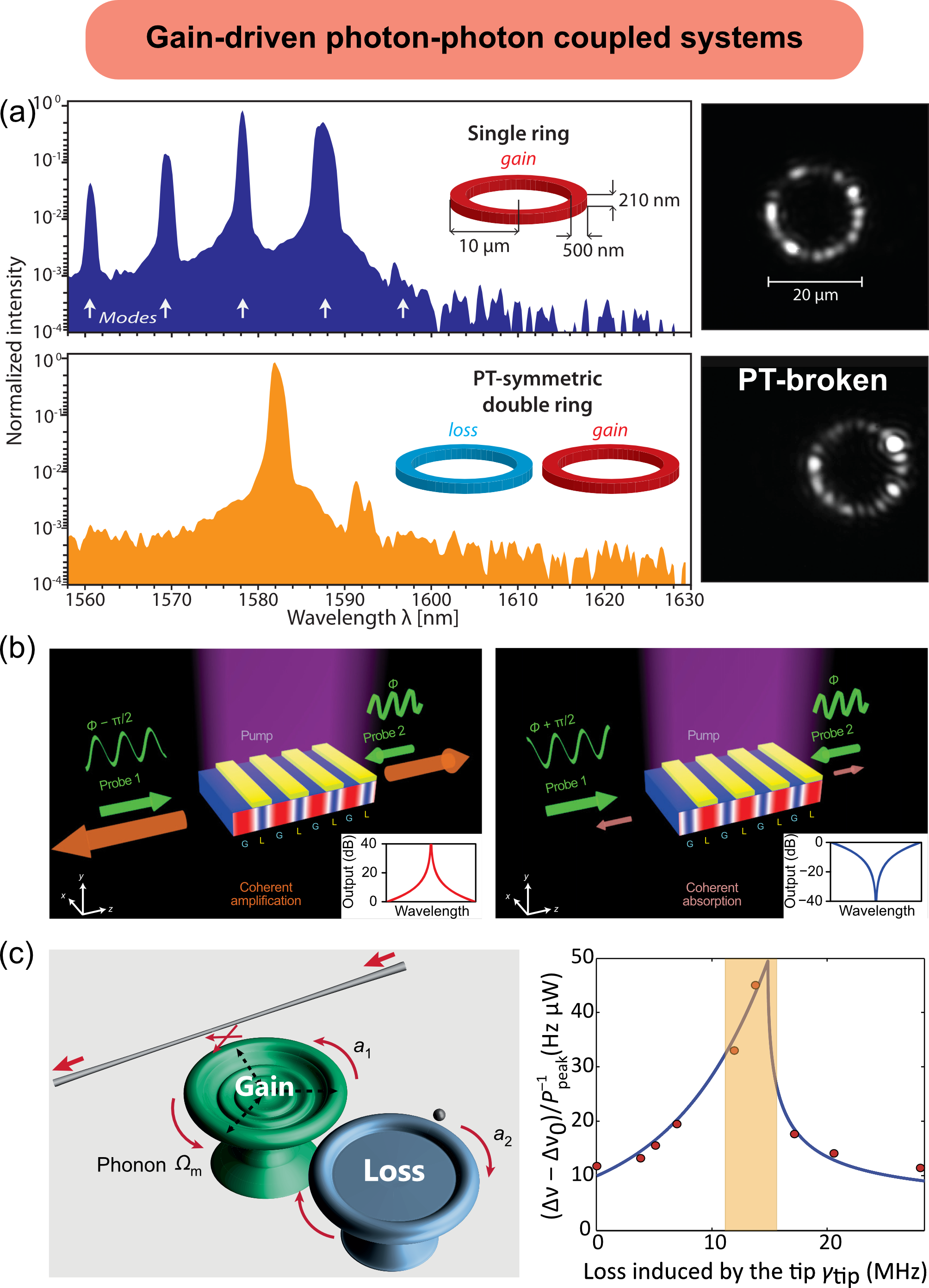}
		\caption{(a) Output spectra of a single ring laser and a PT-symmetric double ring\cite{single}, showing spatial emission intensity distributions at right. (b) Lasing and anti-lasing in a PT-symmetric laser cavity\cite{wong2016lasing}, achieved by changing the phase difference between incident pumping signals. The medium inside the cavity is periodically patterned as gain and loss regions. (c) Gain-loss coupled microcavities\cite{zhang2018phonon}, where gain-driven microcavity generates phonons through optomechanical coupling. The phonon linewidth reaches a maximum at the EP, marked by the orange shaded area. (a) Reproduced with permission from Ref.~\onlinecite{single}. Reprinted with permission from AAAS. (b) Reproduced with permission from Wong et al., Nat. Photon. 10, 796–801 (2016). Copyright 2016, Springer Nature. (c) Reproduced with permission from Zhang et al., Nat. Photon. 12, 479–484 (2018). Copyright 2018, Springer Nature.}
		\label{fig5}
    \end{figure}
    Lasers, being optical gain-driven oscillators, are well-suited for studying PT-symmetry in optics\cite{sciencePToptics,ozdemir2019parity,9072282,Wang23}. The gain-loss coupled laser system consists of a laser mode and a lossy mode, where the lossy mode can be realized through a semiconductor cavity without the gain medium. Then, two cavities can be dimensionally identical and resonate at the same frequency. Benefiting from advanced microfabrication, the gain and loss of the laser medium can be precisely controlled\cite{pumplaserPT,revival}, allowing these coupled laser systems to be designed with PT-symmetry\cite{OpticalPT,doppler2016dynamically}. Investigations into the coupled dynamics between optical modes characterized by gain and loss have revealed intriguing phenomena.

    PT-symmetry assigns an unconventional role to loss in coupled laser systems. In such systems, increasing loss can either suppress or activate emission \cite{revival,feng2014single,single,singleTrans}, challenging the conventional wisdom of single oscillators where loss typically suppresses gain-driven oscillations. This counterintuitive phenomenon has found practical applications. For instance, as shown in Fig. \ref{fig5}(a), it enables achieving single-mode emission in coupled ring lasers \cite{single} by suppressing sidebands through strong coupling while maintaining emission of the laser peak through weak coupling. Consequently, high-power single-mode lasers can be attained \cite{cseker2023single}. Moreover, this mechanism has been employed in optoelectronic oscillators to achieve low phase noise microwave emission \cite{optoe,liu2018observation,zhang2020parity,Dai23}.

    PT-symmetry also enables innovative laser designs. In weakly coupled systems, eigenstates exhibit two complex eigenvalues, indicating lasing and absorbing\cite{twomodelase,newlaser,newlaserstripe}. As shown in Fig. \ref{fig5}(b), these eigenstates have been experimentally realized in a single cavity using patterned medium\cite{wong2016lasing}, where the medium is engineered to be spatially periodic with precisely balanced gain and loss. The two eigenstates can be selectively activated by monochromatic pumping with two incident signals. By varying the phase difference between incident waves, distinct eigenstates can be selectively excited, allowing for either lasing or absorbing within the same laser cavity. This innovation holds promise for integrating the optical amplifier and attenuator into a single device.

    The emission linewidth of PT-symmetric laser systems has attracted significant interest, as it reflects the output signal quality. Theoretical reports predict a broadened linewidth at the EP\cite{GwangsuPfactor}, attributed to changes in the Petermann factor\cite{Petermann,HenningPfactor,GeneralPfactor,abPfactor}, which arises from the non-Hermiticity of open systems. As shown in Fig. \ref{fig5}(c), experimental verification has been performed on PT-symmetric optical cavities\cite{zhang2018phonon}. The gain cavity is optomechanically coupled with acoustic oscillation\cite{optoMech}, allowing the linewidth of the coupled system to be indirectly monitored through the linewidth of the acoustic emission. By precisely tuning the loss of the loss cavity, a significant broadening of the emission linewidth is observed around the EP. 
    
    PT-symmetric lasers constitute a substantial portion of the broader field of non-Hermitian optics. Beyond laser systems, coupled waveguide systems also serve as a platform for PT symmetry in optics, where the temporal dynamics are replaced by spatial longitudinal propagation along the waveguides\cite{guo2009prl,ruter2010observation,Miri2012}. For further related reading in non-Hermitian optics, we recommend the following review resources: Ref.~\onlinecite{ozdemir2019parity},~\onlinecite{Wang23},~\onlinecite{sciencePToptics}, and~\onlinecite{nsrPTphoton}.

    \subsubsection{PT-symmetric magnonic systems}
    \begin{figure}[tbp]
		\centering
		\includegraphics{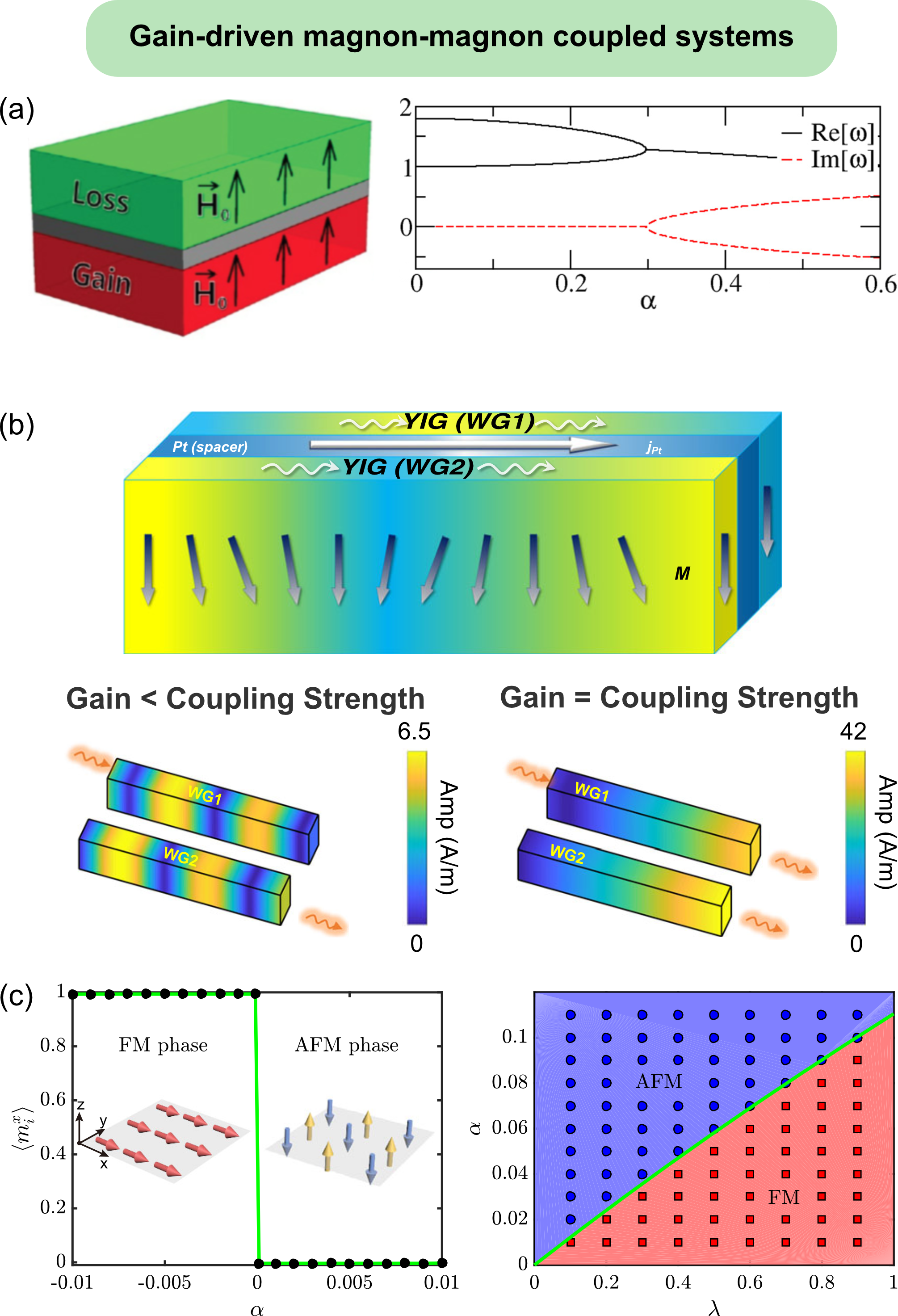}
		\caption{(a) PT-symmetric magnetic films in sandwich model\cite{PTmagnon}, where the coupled ferromagnetic resonance frequency shows an EP by changing the value of Gilbert damping coefficient $\alpha$. (b) Schematic of a magnetic sandwich structure\cite{wang2020steering}, where current in the metal layer generates opposite-direction spin currents in two parallel magnetic layers, serving as magnon waveguides with adjustable input-output relations. (c) Gain-induced collective phase transition\cite{AFMPT}. Left panel: Changing the damping coefficient from negative to positive in a ferromagnetic film induces collective dynamics, leading to a phase transition to the antiferromagnetic phase. Right panel: Phase diagram for damping coefficient $\alpha$ and interlayer coupling strength $\lambda$ in a PT-symmetric magnetic model. (a) Adapted with permission from Ref.~\onlinecite{PTmagnon}. Copyrighted by the American Physical Society. (b) Reproduced with permission from Wang et al., Nat. Commun. 11, 5663 (2020). Copyright 2020, licensed under a Creative Commons Attribution (CC BY) License. (c) Adapted with permission from Ref.~\onlinecite{AFMPT}. Copyright 2018 licensed under a Creative Commons Attribution (CC BY) License.}
		\label{fig6}
    \end{figure}
    Non-Hermitian dynamics is also studied in magnonic systems. The original work by B. Heinrich and his collaborators on coupled magnetic bilayers provided a valid experimental and theoretical platform for this research\cite{rkky,pumping}. J. M. Lee et al.\cite{PTmagnon} theoretically proposed it as a PT-symmetric magnonic system, exhibited as a magnetic sandwich model shown in Fig. \ref{fig6}(a). In this model, the upper magnetic film is assigned as a macroscopic gain-driven magnetization induced by negative Gilbert damping, while the bottom film experiences equal loss due to positive Gilbert damping. These two macroscopic magnetizations are assigned as coherently coupled through either exchange or dipole-dipole interactions. By tuning the gain coefficient, the magnetic resonance of this system exhibits an EP. This simple structure provides a concise model for exploring non-Hermitian physics, attracting significant interest in the magnonics community. Although numerous methods exist to explore non-Hermitian magnonics\cite{ESmagnon,APTzhao,twospinDissiC,YI2020,stoChain,NHSEyutao,twospinPT,wittrock2024non}, studies on non-Hermitian magnetic system are mostly inspired by this model\cite{Yunshan2019,sciadvPT,enPT,epWGmagnon,transMagnonPT,Deng2023,FloquetPT,diodePT}.

    This PT-symmetric model sustains intriguing physics properties around the EP, potential for magnonic devices. A simulation of a sandwich structure composed of yttrium iron garnet (YIG) and platinum (Pt) layers exhibits its potential as a magnonic waveguide\cite{wang2020steering}, as shown in Fig. \ref{fig6}(b). The Pt interlayer, characterized by a strong inverse spin Hall effect\cite{SHEs}, generates a spin current perpendicular to the film plane when a current is applied. This injected spin current imparts opposing spin torques to the two magnetic layers, respectively enhancing the damping and spin torques. Consequently, gain and loss are introduced to the magnons in the two layers. Serving as a magnonic waveguide, the input-output relation of the magnetic layers exhibits two features: first, the magnons can be amplified through the device; second, magnon propagation can be nonreciprocal around the condition of EP. Since the gain and loss coefficients are determined by the current flowing through the interlayer, this device's nonreciprocity can be manipulated through the current.

    Attracted by the potential magnonic applications featured by the EP, investigations, including thermal excitation\cite{diodePT} and Floquet modulation\cite{FloquetPT}, have been performed on the sandwich structure and even extended to multilayer structures\cite{enPT}. Experiments using Brillouin light scattering spectroscopy have demonstrated the feasibility of PT-symmetric magnetic waveguides on dipole-dipole coupled YIG stripes\cite{epWGmagnon}, indicating a promising future for PT-symmetric magnonic devices.

    PT-symmetric magnonic systems also exhibit unique characteristics, featuring fascinating collective dynamics caused by interactions between spins\cite{AFMPT}. Recent reports predict that a ferromagnet with gain (loss) is equivalent to an antiferromagnet with an equal value of loss (gain)\cite{AFMPT}, as illustrated in Fig. \ref{fig6}(c). This equivalence arises from the collective dynamics of spins, opening new possibilities for designing non-Hermitian magnonic devices. Simultaneously, this finding, consistent with studies including experiments in spin injection\cite{LinPRL2016} and theoretical work on spin-torque oscillators\cite{storquePT}, highlights the distinction between collective gain-driven oscillations and single gain-driven oscillations. They suggest that gain-loss coupled magnonic systems might be more sophisticated, and potentially associated with material phase transitions.

    PT-symmetric magnonic systems have captured significant interest within the magnonics community, inspiring the rapid development of non-Hermitian magnonics. For further related reading, we recommend the following review resources: Ref.~\onlinecite{Hurst2022},~\onlinecite{YUAN20221},~\onlinecite{YU20241}, and~\onlinecite{ZhangAEM}.
    
\subsection{Nonlinear gain-loss coupled systems}\label{sec3C}
    The gain-loss coupling has also been investigated in nonlinear systems, where nonlinearity emerges from the amplified oscillation amplitude. Current research on nonlinear gain-loss systems primarily focuses on steady states due to their potential applications in wave transfer and wave generation. These include wireless power transfer systems, exciton-polariton systems, cavity magnonic systems, and magnetoacoustic systems.

    \subsubsection{PT-symmetric wireless power transfer systems}
    \begin{figure}[t]
		\centering
		\includegraphics{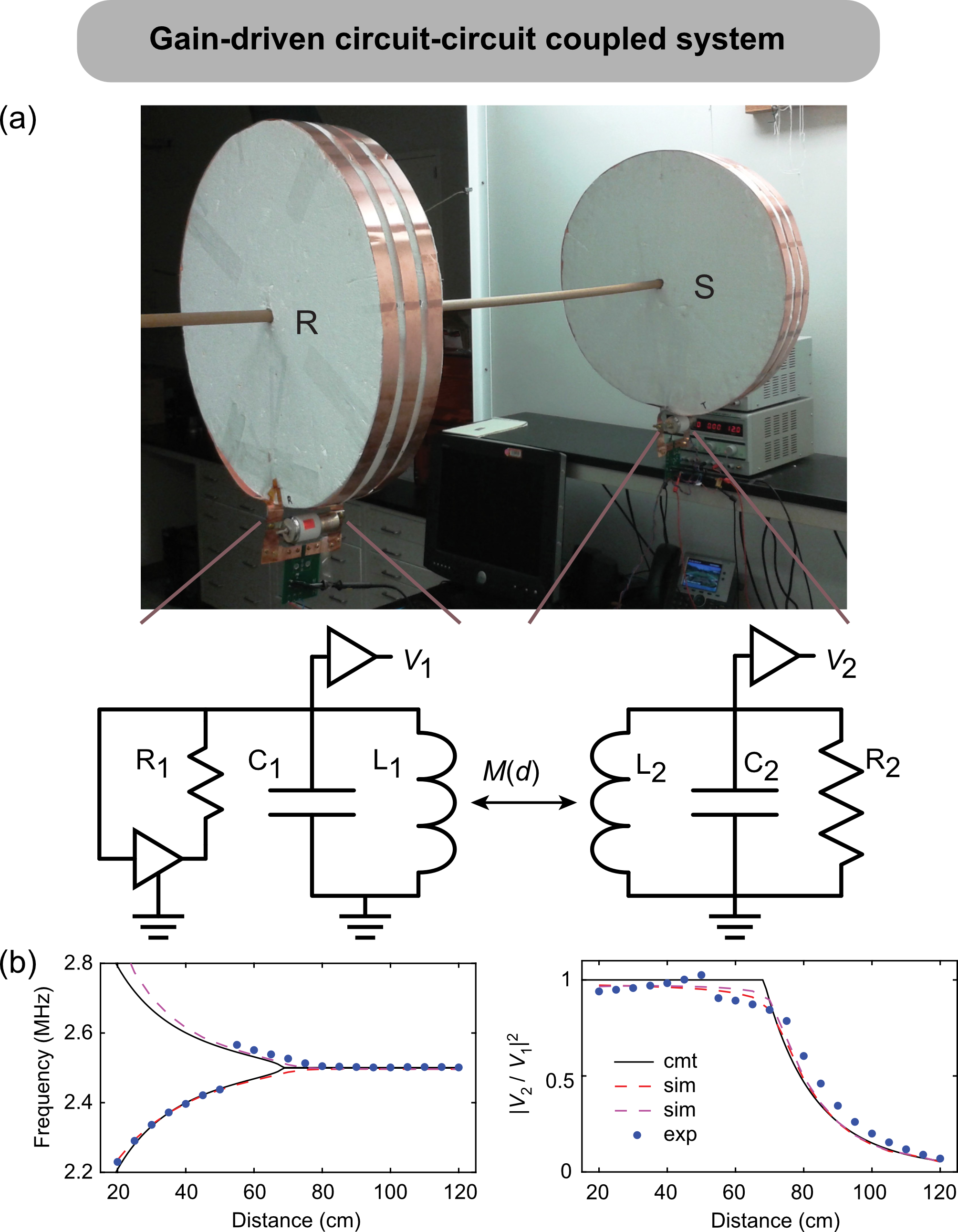}
		\caption{(a) Schematic of WPT setup\cite{assawaworrarit2017robust}, where two LRC circuits are coupled through the mutual inductance. Energy is transferred from the amplifier-embedded LRC circuit to the damped LRC circuit. (b) Corresponding voltage ratio and frequency of coupled-mode theory (cmt), circuit simulation (sim), and experimental (exp) results are given. Reproduced with permission from Assawaworrarit et al., Nature 546, 387–390(2017). Copyright 2017, Springer Nature.}
		\label{fig4}
    \end{figure}

    An electronic oscillator and a damped resonator can be respectively modeled as the energy source and energy receiver in a PT-symmetric wireless power transfer (WPT) system. This coupling enables energy transfer via electromagnetic waves between two circuits without physical contact, facilitating the convenient deployment of electronic devices in industry\cite{IEEEWPT,droneWPT,song2021wireless}. The energy transfer device, schematically shown in Fig. \ref{fig4}(a), was first proposed by S. Assawaworrarit et al\cite{assawaworrarit2017robust}. An electronic oscillator is coupled with a damped LRC circuit through the mutual inductance of the inductors. It allows the coupling strength to be adjusted by changing the distance between the circuits.

    This type of WPT system is typically depicted by a PT-symmetric Hamiltonian. Under the zero-detuning approximation, PT-symmetric circuit theory aligns well with experimental results. A key feature is that the amplitude ratio, representing power transfer efficiency, remains unitary in the strong coupling regime before the EP. Thus, optimizing a WPT setup involves engineering the PT-symmetric system by adjusting the EP's position\cite{zhang2023robustness}.
    
    However, a distinct difference observed in experiments is the presence of solely one oscillation frequency in the gain-loss circuit, contrasting with the theoretical prediction of two eigenmodes. This discrepancy arises from ignoring van der Pol nonlinearity. Recent reports indicate that nonlinearity, including van der Pol and Duffing nonlinearities, may influence power transfer stability and efficiency\cite{bistableWPPT}.
	
    Incorporating a third mode\cite{genralPT} into the gain-loss coupled circuit can further optimize WPT systems in terms of stability and transfer distance. The third mode acts as an intermediary between the gain-driven source mode and the lossy receiver mode, which can be a gain-driven mode\cite{frobust}, a zero-damping mode\cite{frobust}, a lossy mode\cite{horderEP,nsrAPT}, an oscillation network\cite{chainWPT,sciadvWPT,Wu_2024}, or even metamaterials\cite{wang2023metamaterial}. Although these systems are beyond the regime of the PT-symmetric model, the focus remains on maintaining a stable and efficient energy transfer process by strategically manipulating EPs\cite{horderEP}. 
    
    PT-symmetric WPT systems have potential applications in industry, such as powering electric vehicles, biomedical implants, and portable devices. For further related reading, we recommend the following review resources: Ref.~\onlinecite{song2021wireless} and Ref.~\onlinecite{gao2023two}.
    
    \subsubsection{Exciton-polariton system}
    \begin{figure}[tbp]
		\centering
		\includegraphics{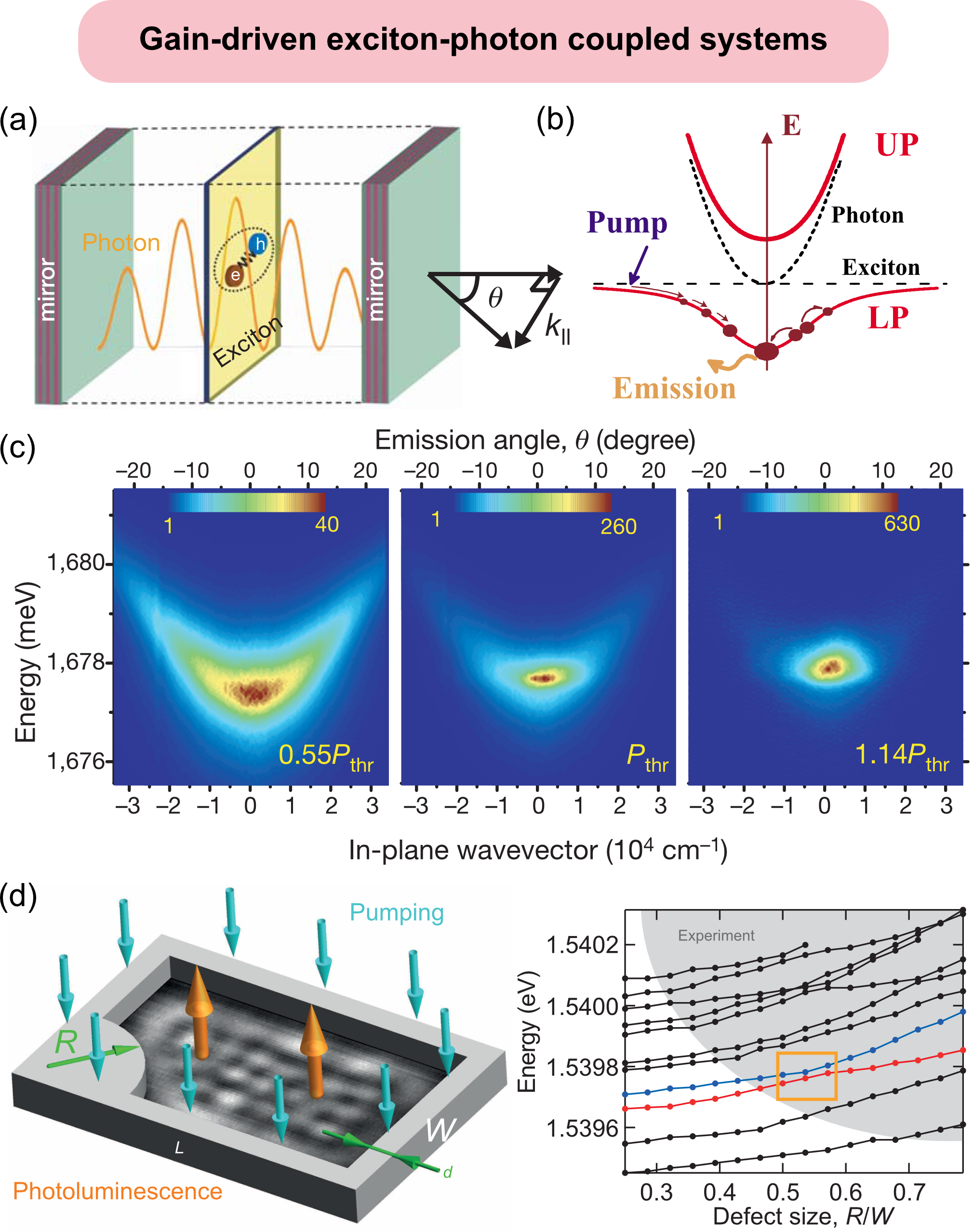}
		\caption{(a) Optical microcavity photon coupling with exciton results the exciton polariton\cite{kasprzak2006bose}. (b) Pumping and photon leakage of strong coupled system\cite{DengPolaritonBEC}. Upper polariton (UP) and lower polariton (LP) caused by strong coupling, where polariton is pumped at the exciton's enegy level of resulting the gain of exciton polariton. The photon leakage leads to the loss of exciton polariton. (c) Condensation of leakage photon's energy distribution at different pumping power\cite{kasprzak2006bose}. (d) Schematic experimental setup of an exciton-polariton billiard\cite{gao2015observation}, and the measured energy spectrum. Defect size is adjusted, realizing two near-degenerate modes. (a,c) Reproduced with permission from Kasprzak et al., Nature 443, 409–414 (2006). Copyright 2006, Springer Nature. (b) Adapted with permission from Ref.~\onlinecite{DengPolaritonBEC}. Copyrighted by the American Physical Society. (d) Reproduced with permission from Gao et al., Nature 526, 554–558 (2015). Copyright 2015, Springer Nature.}
		\label{fig7}
    \end{figure}
    The exciton-polariton system is usually a hybrid system where cavity photons interact with excitons within semiconductor materials\cite{DengPolaritonBEC,byrnes2014exciton}. As illustrated in Fig. \ref{fig7}(a), the semiconductor material is fabricated into an optical microcavity with quantum wells\cite{kasprzak2006bose}, housing paired electrons and positive holes. These pairs, known as excitons, are characterized by the interaction between electrons and holes without merging\cite{nonlinearExciton}. Excitons can strongly interact with cavity photons, forming coupled hybrid states called exciton-polaritons. This system is famous for the luminescence of the exciton-polariton condensate\cite{polaritonlaser,micropillarPolariton,solidPolaLaser}, making it a promising optical source. Structurally, the exciton-polariton BEC system, also named as the polariton laser, resembles a semiconductor laser, as both contain a cavity and an internal material. However, unlike the laser's gain medium that amplifies light, the exciton material provides a second mode, excitons, which couple with cavity photons.

    Non-equilibrium Bose-Einstein condensation can be observed in pumped exciton-polariton systems.  Unlike traditional cold-atom BECs, realized through thermalization in conservative systems, the exciton-polariton condensates are formed in an open system driven by external pumping\cite{ddphysics}. As depicted in Fig. \ref{fig7}(b), the strong coupling between excitons and photons results in two distinct polariton branches\cite{kasprzak2006bose,Wouters,byrnes2014exciton}, lower polaritons (LP) and upper polaritons (UP). The energy of LPs dissipates as photon emission\cite{PolaritonDicke}, leading to loss. To balance this loss, an incoherent pump is applied, achieving gain through a nonlinear process called stimulated scattering\cite{Goldstone} which is similar to a parametric pumping process. When gain and loss are balanced, the system reaches a steady state. Once the density of polaritons exceeds a certain threshold, this steady state condenses, allowing a large population of polaritons to occupy the ground energy level, forming the non-equilibrium polariton BEC. 
    
    Exciton-polariton condensates are nonlinear states. As shown in Fig. \ref{fig7}(c), the condensation is experimentally evidenced by the reduced emission angle of photons\cite{kasprzak2006bose}. Since the angle corresponds to the momentum of the polaritons, the reduced angle indicates that the polaritons are being condensed to the ground energy with zero momentum. The condensation results in two features. First, all the polaritons occupy the ground energy level, described by the same wave function. Second, the polaritons occupying the same energy level correlate with each other, leading to strong nonlinearity. Based on these features, the macroscopic dynamics of condensed polaritons are described by a Schrodinger-type equation with a nonlinear term, known as the Gross–Pitaevskii equation\cite{Pethick_Smith_2001}. Thus, for non-equilibrium exciton polariton, the theory describing the evolution of condensate incorporates gain, loss, and nonlinearity\cite{Wouters,NHPBEC}.

    EPs are observed in exciton-polariton systems. As shown in Fig. \ref{fig7}(d), the exciton-polariton condensate is placed within a billiard potential well designed with a defect\cite{gao2015observation}. Confined by the billiard, the state of polariton condensate shows multiple energy levels. By adjusting the size of the defect, the energy levels tend to coalesce, reflecting the degeneracy of real eigenvalues of the exciton-polariton system. This observation further demonstrates the non-Hermitian nature of the exciton-polariton system. Based on such a nonlinear gain-loss coupled system, collective phenomena including phase transitions and fluctuations are becoming prominent research topics\cite{NHPBEC,fluctuationsPolaritonBEC,NonReciprocalDicke,sieberer2023universality}.

    For further related reading, we recommend the following review resources: Ref. ~\onlinecite{DengPolaritonBEC}, ~\onlinecite{byrnes2014exciton}, ~\onlinecite{ZHANG2022100399}, ~\onlinecite{BallariniDeLiberato}, and ~\onlinecite{bloch2022non}.

    \subsubsection{Cavity-magnonic system}
    Cavity-magnonic systems, a type of light-matter coupled system, are composed of interacting microwave cavity photons and magnons\cite{Harder2021,ZARERAMESHTI20221}. The systems are characterized by strong coupling strength due to the large amount spins inherent to magnons\cite{strongMP,Huebl2013,xufeng2014}, making them promising as quantum transducers\cite{clerk2020hybrid,hybridTransucer}. Recently, the gain-loss cavity-magnonic systems have been investigated, showing potential as microwave and magnon sources.

\begin{figure}[tbp]
		\centering
		\includegraphics{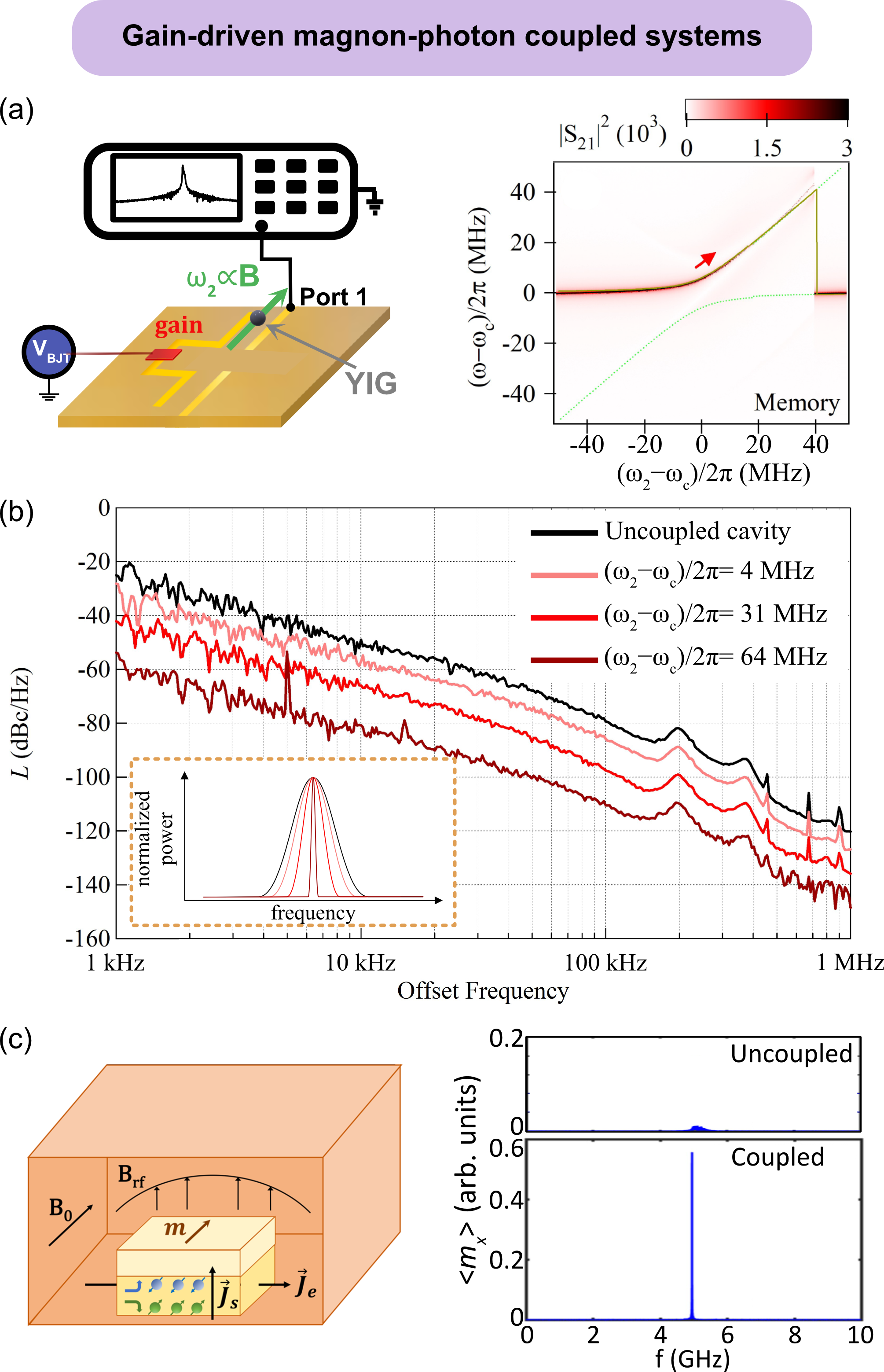}
		\caption{(a) Schematic and frequency dispersion of a gain-loss cavity-magnonic system based on electronic oscillation\cite{GDP}. (b) Phase noise for the uncoupled cavity and coupled cavity at different frequency detunings\cite{kim2024low}. Inset shows the schematical evolution of emission peaks for different phase noises. (c) Schematic of a gain-loss cavity-magnonic system based on spin-torque oscillation\cite{STOmaser}, showing spectra of precessing magnetization at uncoupled and coupled conditions. (a) Adapted with permission from Ref.~\onlinecite{GDP}. Copyrighted by the American Physical Society. (b) Reproduced with permission from Kim et al., Appl. Phys. Lett. 124 (2024). Copyright 2024, AIP Publishing. (c) Adapted with permission from Ref.~\onlinecite{STOmaser}. Copyrighted by the American Physical Society.}
		\label{fig8}
    \end{figure}

    Gain can be implemented via electronic microwave oscillation\cite{GDP} or spin-torque oscillation\cite{STOmaser}. These studies are based on the nonlinear paradigm, considering the van der Pol nonlinearity of gain-driven oscillations. In the system where a magnetic sphere couples with an electronic oscillator, as shown in Fig. \ref{fig8}(a), the van der Pol nonlinearity originates from the voltage cap of the electronic amplifier\cite{GDP}. In the system with a spin-torque oscillator inside the microwave cavity, as shown in Fig. \ref{fig8}(c), the van der Pol nonlinearity originates from the large-angle precession of magnetization\cite{STOmaser}.

    Although these systems utilize different gain-driven oscillation mechanisms, they exhibit similar features. First, as shown in Fig. \ref{fig8}(a), only one coupled mode is observable in the frequency dispersion\cite{GDP}. As mentioned in the theoretical section, this property is due to van der Pol nonlinearity, which results from the strong amplitude of the eigenmode close to the oscillator's frequency. Second, the emission peak of the polariton is significantly sharpened in the coupled condition\cite{kim2024low}. This phenomenon is systematically investigated through phase noise distribution, as shown in Fig. \ref{fig8}(b). The reduced emission linewidth, indicated by low noise, demonstrates that the thermal dynamics of cavity-magnonic polariton undergo a convergence process. From an engineering perspective, the Lesson's equation\cite{1446612,rubiola2005leeson}, which considers the oscillators' quality factors as the key determinant of phase noise, cannot describe this novel phase noise reduction process without directly altering the quality factor of the magnet and microwave cavity.

    Other methods can also realize the gain-loss cavity-magnonic system. The original work, conducted in 2003 by Eliyahu and Maleki\cite{1210597}, preceded the studies mentioned in this section. Their system comprised an optoelectronic oscillator and a lossy YIG film, resulting in a frequency-tunable microwave oscillator. Theoretically, microwave parametric pumping\cite{Mukhopadhyay,Huang22} and laser-induced magneto-optical interactions\cite{NGD} also show high feasibility for achieving gain-driven magnetic oscillation. These proposals are based on mature electronic and optical techniques\cite{7289369,10117091,xiong2024magnon}, suggesting a promising future for the gain-loss cavity-magnonic system.
    
    \subsubsection{Magnetoacoustic system}
    \begin{figure}[tbp]
		\centering
		\includegraphics{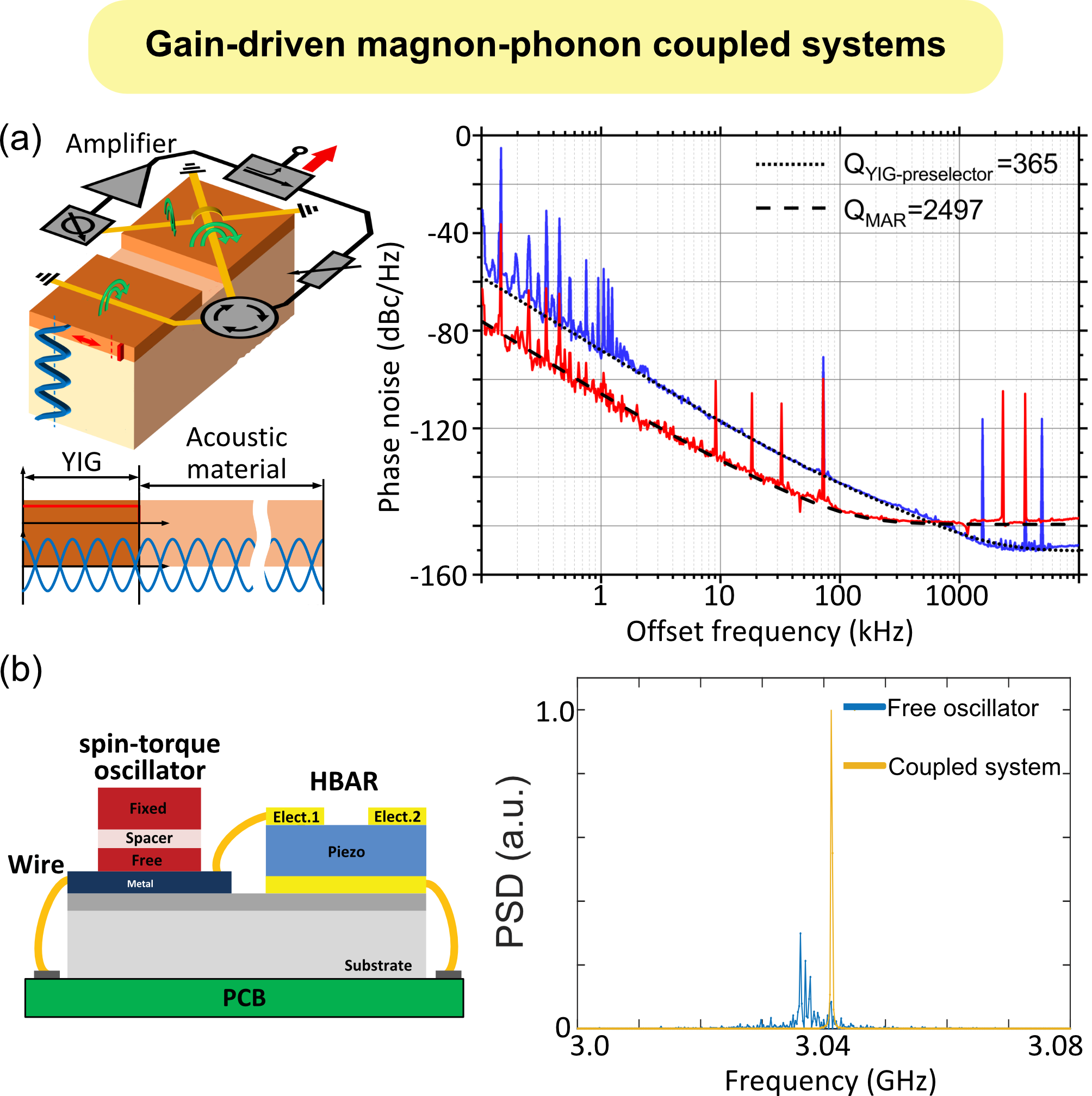}A
		\caption{(a) Schematic of a magnetoacoustic oscillator and its phase noise profile\cite{maoscillator}, comparing magnetic (red) and magnetoacoustic (blue) oscillators. (b) Schematic of a spin-torque oscillator coupled with an HBAR\cite{torunbalci2018magneto}. The oscillator's spectral at the coupled condition shows a single sharp peak and an enhancement of emission power. (a) Adapted with permission from Ref.~\onlinecite{maoscillator}. Copyrighted by the American Physical Society. (b) Reproduced with permission from Torunbalci et al., Sci. Rep. 8, 1119 (2018). Copyright 2018, licensed under a Creative Commons Attribution (CC BY) License.}
		\label{fig9}
    \end{figure}

    Low-damping magnetic or piezoelectric materials are commonly embedded in electronic oscillators to optimize the quality factor, known as magnet or crystal oscillators\cite{marrison1948evolution,matthys1983crystal,ishak1988magnetostatic,rumyantsev2019amplitude}. This approach has inspired engineers to integrate magnet oscillators with high-quality acoustic resonance, aiming to utilize the acoustic resonance peak as a band-pass filter for noise signals, thereby achieving a sharp emission linewidth. This concept has facilitated experiments on gain-loss magnetoacoustic systems. The coupling between magnetization oscillation and acoustic resonance can be realized in two ways: (1) through magnetostriction, for example, when magnets are in direct contact with acoustic materials\cite{1535919,10144384}, (2) via wire connection through microwave voltage or current due to the piezoelectricity of certain acoustic materials\cite{torunbalci2018magneto}.

    These coupled systems function as gain-loss hybrid systems, exhibiting characteristics similar to gain-loss cavity-magnonic systems. Experimentally, the magnetoacoustic oscillator is realized using a double-layer film\cite{maoscillator}, as shown in Fig. \ref{fig9}(a). The magnetic film layer is embedded in a feedback loop, inducing magnetization oscillation. This oscillation couples with the acoustic resonance of the bottom layer through magnetostriction. The measured phase noise curves demonstrate an enhanced quality factor in the hybrid oscillator compared to the magnet oscillator alone. Theoretical studies, based on spin-torque oscillators (Fig. \ref{fig9}(b)), also predict a sharper emission peak compared to the uncoupled oscillator\cite{torunbalci2018magneto}.

    \begin{figure}[tbp]
		\centering
		\includegraphics{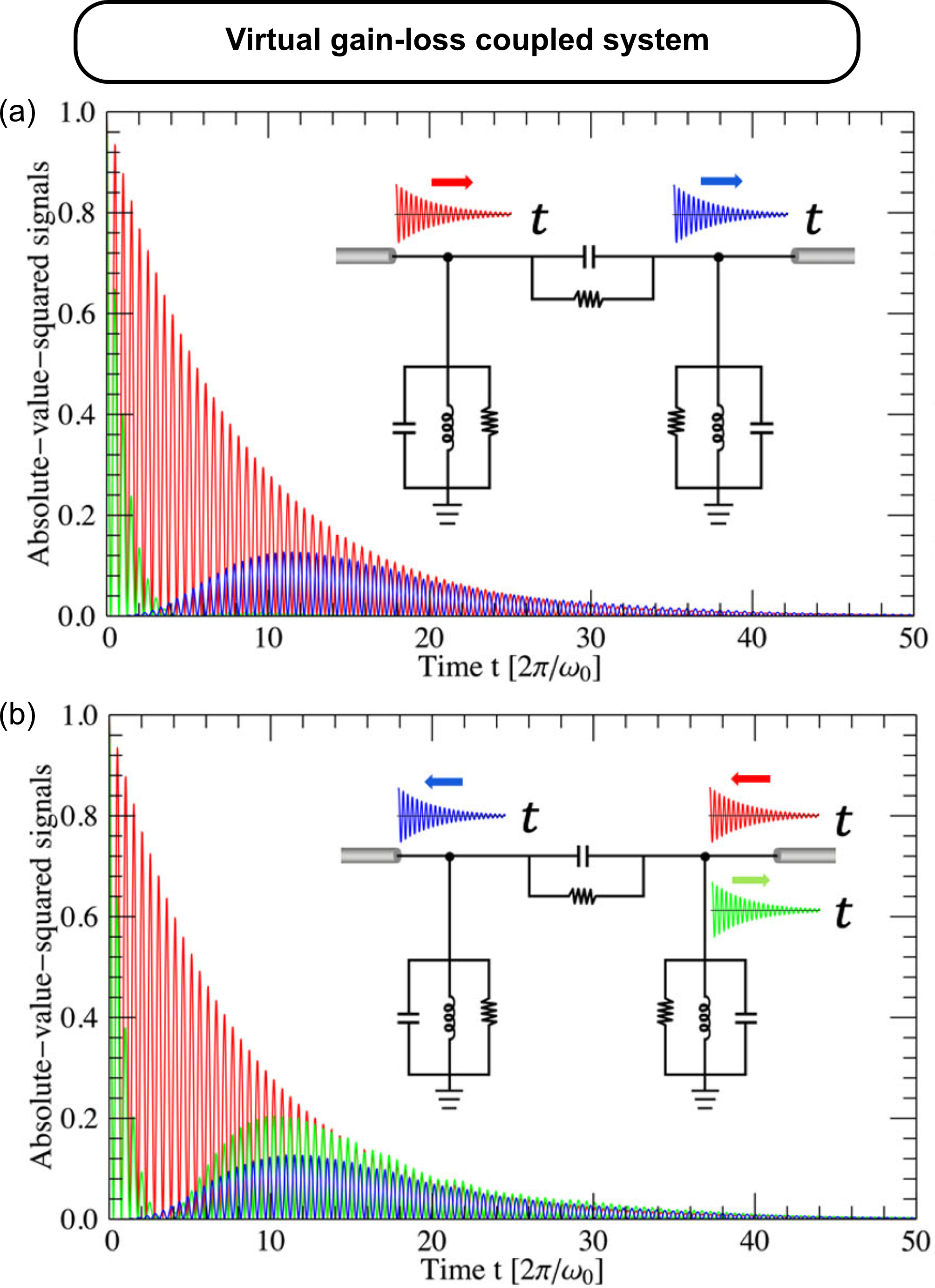}
		\caption{Theoretical demonstration of a virtual gain-loss coupled circuit. Two passive resonators are coupled through reactance, with the virtual gain-driven mode realized through complex frequency excitation, while the other passive circuit resonator provides the lossy mode. Time-dependent complex frequency excitation is applied from (a) the left port and (b) the right port. In both cases, the incident (red), reflected (green), and transmitted signals (blue) are plotted. Adapted with permission from Ref.~\onlinecite{VPTS2020}. Copyrighted by the American Physical Society.}
		\label{fig10}
    \end{figure}
    
    Compared to cavity-magnonic systems, magnetoacoustic systems show greater potential for on-chip devices. Acoustic resonance can be achieved using a high-overtone bulk acoustic resonator (HBAR)\cite{4787193}, characterized by its micrometre-scale size and gigahertz frequency operation. Thus, without considering the technical difficulties, the integration of HBARs and magnets is relatively feasible for minimized on-chip devices than the centimetre-sized microwave cavities.

\subsection{Virtual gain-loss coupled system}\label{sec3D}
    Traditional gain-driven oscillations based on realistic gain elements, for example, the gain medium of a laser, are often accompanied by several drawbacks, including instabilities, unwanted nonlinearities, and amplified quantum noise due to increased spontaneous emissions. These issues complicate the behavior of coupled systems and can degrade performance and reliability. In contrast, virtual gain bypasses these problems by mimicking the behavior of active media through complex frequency excitation, allowing control over scattering properties without the drawbacks associated with material gain.

    PT symmetry can emerge in virtual gain-loss coupled systems. As introduced in Ref.~\onlinecite{VPTS2020}, virtual gain-loss coupling is achieved by applying exponentially decaying signals to one of the damped LRC circuits, while the other remains passive. The notable phenomenon enabled by this approach is asymmetric transmission resonance (ATR)\cite{VPTS2020,2012GeLi}, which is characterized by unidirectional invisibility as shown in Fig. \ref{fig10}. Traditionally, achieving ATR in optics requires a gain medium. However, with complex frequency excitation, ATR can be realized without the need for material gain, providing a novel and efficient route to this effect\cite{2024Trivedi}.

\section{Outlook}\label{sec4}
    The discussion on various systems reveals intriguing observations and innovative applications, spanning from classical to quantum systems. In this section, we propose three directions that might attract interest in the context of general physics.
    
    \subsection{Condensation in gain-loss coupled systems}
    Serving as an open system, the condensation of gain-loss coupled systems can not be depicted by the thermodynamic equilibrium of a conservative model\cite{Wouters,sieberer2023universality}. Theoretical efforts have been devoted to investigating open quantum systems, such as the exciton-polariton system. However, by examining the noise distribution of emission spectra, convergence phenomena (explicitly shown as Fig. \ref{fig8}(b) and \ref{fig9}(a)) are also found in classical systems. Features such as high emission power, coherence, and narrow linewidth have been reported. Moving forward, further experimental investigation and theoretical insights are essential to fully understand and exploit the general statistical condensation phenomena in gain-loss coupled systems. These advancements could inspire the understanding of non-Hermitian physics and give rise to advanced wave sources.
    
    \subsection{Non-reciprocity in gain-loss coupled systems}
    In this article, we have discussed non-reciprocity caused by EPs in PT-symmetric systems. However, non-reciprocity is also a well-known phenomenon in nonlinear physics. Since both nonlinearity and EPs can be observed in nonlinear gain-loss coupled models, we anticipate more fruitful non-reciprocal phenomena in gain-loss coupled systems. A recent experiment, done by Zhang et al.\cite{Zhang} on a nonlinear circuit platform, has shown an intensive energy transfer from a gain-driven oscillator to a lossy oscillator. Similar theoretical predictions have been reported in the quantum battery system, recognized as the non-reciprocal phenomena caused by the breaking of time-reversal symmetry\cite{nrepBattery}. Further experiments and theoretical investigation are expected on the advanced platforms.

    \subsection{Dissipative coupling in gain-loss coupled systems}
    Current research on gain-loss coupled systems is primarily focused on coherent coupling. However, in dissipatively coupled systems, such as dissipative coupled cavity-magnonic systems, two modes can indirectly couple through oscillatory mediation (such as a damped mode or travelling wave), resulting in level attraction\cite{wangdc,YUdc,DCreview}. Since the mediating oscillation might be influenced by gain, it is significant to investigate whether dissipative coupling can still be induced by mediation in gain-loss coupled systems. As dissipative coupling is emerging as an attractive mechanism for synchronization, non-reciprocity\cite{Yuan2023,ZYWang2023}, and lasing within the condensed matter physics community, studies on dissipative coupling in these new systems could potentially unveil new physics and insights.
    
\section{Conclusion}\label{sec5}
    In our review of various gain-loss coupled systems, we categorize these systems into two main models: linear and nonlinear gain-loss coupled models. Gain, often generated from non-resonant excitation, activates intensive oscillations that result in nonlinearity. Systems focusing on phase transitions caused by gain are described by the linear model, where nonlinearity is neglected. In contrast, systems focusing on intensive oscillations incorporate nonlinearity into the coupled model, resulting in the nonlinear gain-loss model. Compared to linear gain-loss coupled systems, systems described as nonlinear models exhibit additional effects such as general condensation phenomena and self-selection.

    Based on these gain-loss coupled models, we propose three future research directions: condensation, non-reciprocity, and dissipative coupling. These aspects relate to steady-state behaviors, phase transitions, nonlinearity, and coupling mechanisms. The field of gain-loss coupled systems is vast and promising, offering exciting opportunities for exploring new physics and advancing practical applications.

\begin{acknowledgments}
    We are grateful for the communication with Michael Cottam, Justin Hou, Benjamin Jungfleisch, Weiwei Lin, Luqiao Liu, Mohammad-Ali Miri, Jie Qian, Jiang Xiao, John Q, Xiao, Peng Yan, Ying Yang, Weichao Yu, and Huaiyan Yuan. This work has been funded by NSERC Discovery Grants, NSERC Discovery Accelerator Supplements, Innovation Proof-of Concept Grant of Research Manitoba, and Faculty of Science Research Innovation and Commercialization Grant of University of Manitoba (C.-M. H.). C. Z. is supported by the China Scholarship Council (Grant No.CSC202106180012).
\end{acknowledgments}

\appendix
\section{Derivation of first-order van der Pol equation} \label{AppvdP}
    In this section, we derive the first-order van der Pol equation using the averaging method. The standard form of the van der Pol oscillator is expressed as a second-order differential equation,
    \begin{equation}
        \frac{d^{2}a}{dt^{2}}+\omega_{0}=(2g-\gamma_{0} a^{2})\frac{da}{dt},
        \label{AppEq2ndvdP}
    \end{equation}
    where $\omega_0$ denotes the natural frequency of oscillation, $g$ represents system gain, and $\gamma_0$ characterizes nonlinear damping. For a harmonic oscillator, the solution takes the sinusoidal form $a_s = a_0 \cos(\omega_0 t + \phi)$, where $a_0$ and $\phi$ correspond to the amplitude and phase, respectively. To apply the averaging method to Eq. \ref{AppEq2ndvdP}, we introduce the auxiliary function $h(a)$\cite{strogatz2015nonlinear}
    \begin{equation}
        h=(2-\gamma_{0}/g a^{2})\frac{da}{dt}.    
    \end{equation}
    This function can be approximated by
    \begin{equation}
        h(a_{0},\tau)=(2-\gamma_{0}/g a^{2})a_{0}\omega_{0}cos(\tau+\phi),
    \end{equation}
    where $\tau = \omega_0 t$. Consequently, the time evolution of the amplitude and phase can be written as\cite{strogatz2015nonlinear}
    \begin{equation}
        \begin{aligned}
            & \ \frac{da_{0}}{dt}=g[\frac{1}{2\pi\omega_{0}}\int_{0}^{2\pi}sin(\tau-\phi) h\,d\tau], \\
            &\frac{d\phi}{dt}=g[\frac{1}{2\pi\omega_{0}a_{0}}\int_{0}^{2\pi}cos(\tau-\phi) h \,d\tau].
        \end{aligned}
    \end{equation}
    These equations describe the slow-time evolution of the amplitude and phase over long timescales, simplified as
    \begin{equation}
        \frac{da_{0}}{dt}=(g-\frac{\gamma_{0}}{8}a_{0}^{2})a_{0},\ \ \ \frac{d\phi}{dt}=0,
    \end{equation}
    We now express the solution in exponential form using the complex notation: $a = a_0 e^{-i \omega_0 t - \phi}$, equivalent to the sinusoidal solution $a_s = \text{Re}[a]$. The time evolution of $a$ is then
    \begin{equation}
		\frac{da}{dt}=-i\omega_{0}a+(G-\gamma|a|^{2})a.
    \end{equation}
    where $\gamma = \frac{\gamma_0}{8}$. Thus, the second-order van der Pol equation is reformulated as a first-order equation, capturing the dynamics of the system in terms of the complex amplitude $a(t)$.

\section{Derivation of virtual gain} \label{APPvirtualG}
    To illustrate the virtual gain effect, consider the mode evolution in a single resonator with intrinsic loss ${{\kappa }_{L}}$ connected to an excitation port with rate ${{\kappa }_{E}}$. Consider a resonator excited by a signal described by $s(t)={{s}_{0}}{{e}^{-i({{\omega }_{r}}-i{G})t}}={{s}_{0}}{{e}^{-i{{\omega }_{r}}t}}{{e}^{{-G}t}}$, where ${{\omega }_{r}}$ and ${-G}$ are the real and the imaginary parts of the excitation frequency. The governing equation for the mode amplitude $a(t)$ in this open system is $\frac{da}{dt}=-(i{{\omega }_{0}}+{{\kappa }_{L}}+{{\kappa }_{E}})a+\sqrt{2{{\kappa }_{E}}}s(t)$, where ${{\omega }_{0}}$ is the resonant frequency of the resonator, $a(t)$ represents the mode amplitude within the system. Assume a solution for the mode amplitude of the form: $a(t)={{a}_{0}}{{e}^{-i({{\omega }_{r}}-i{G})t}}={{a}_{0}}{{e}^{-i{{\omega }_{r}}t}}{{e}^{{-G}t}}$, where ${{\omega }_{r}}$ is the frequency of oscillation, ${-G}$ represents the decay (${G}>0$) component, aligning with the excitation signal. Substituting the assumed solution into the governing equation yields
    \begin{equation}
        \frac{d{{a}_{m}}}{dt}=-(i{{\omega }_{0}}+\kappa-{G})a_{m}+\sqrt{2{{\kappa }_{E}}}{{s}_{m}}(t),
    \end{equation}
    where $\kappa={{\kappa }_{L}}+{{\kappa }_{E}}$ is the total loss, ${{a}_{m}}(t)={{a}_{0}}{{e}^{-i{{\omega }_{r}}t}}$ and ${{s}_{m}}(t)={{s}_{0}}{{e}^{-i{{\omega }_{r}}t}}$ represent the harmonic part of the mode amplitude and excitation signal. Thus, exciting the system with a complex frequency field effectively introduces the complex frequency as an additional loss term in the system's master equation.

    Virtual gain arises from a decaying excitation signal. In the case $G\leq{{\gamma }_{L}}+{{\kappa }_{E}}$, the effective loss in the system becomes: ${{\kappa }_{\text{eff}}}={{\kappa }_{L}}+{{\kappa }_{E}}-G$, that is the effective loss ${{\kappa }_{\text{eff}}}$ is reduced compared to the total intrinsic loss ${{\kappa }_{L}}+{{\kappa }_{E}}$. This reduction implies that the system behaves as if it has additional gain, thereby increasing the quality factor of the resonator, $Q={{\omega }_{0}}/2{{\kappa }_{eff}}$. Thus, when the decay rate increases to $G>{{\gamma }_{L}}+{{\kappa }_{E}}$, the system is governed by an effective gain coefficient $G_{\text{v}}={{\kappa }_{\text{eff}}}|=G-{{\kappa }_{L}}+{{\kappa }_{E}}$, which represents the so-called virtual gain.
    
\section{Dynamics of gain-loss coupled systems} \label{AppSECH}
    In Sec. \ref{sec2A}, we described a self-oscillator as a gain-driven harmonic oscillator. Building on this, we now provide a simple classical approach to derive Eq. (\ref{Eq2}) and (\ref{Eq3}) for a gain-loss coupled system using the general dynamic equations.
\subsection{Nonlinear dynamics}\label{AppSECnonlinear}
    The coupled dynamics of the gain-driven mode $a$ and loss mode $b$ are governed by the following equations,
    \begin{equation} \label{AppEq1}
        \begin{aligned}
            &\frac{da}{dt}=-i\omega_{1}a+(G-\gamma|a|^{2})a-iJ b. \\
            & \ \ \ \ \ \ \ \frac{db}{dt}=-i\omega_{2}b-\kappa b-iJ a.
        \end{aligned}
    \end{equation}
    where $J$ denotes the coupling strength, $G$ is the gain coefficient,$\gamma$ is the nonlinearity coefficient, and $\kappa$ represents the loss coefficient. These equations can be expressed in matrix form,
    \begin{equation}
        \begin{aligned}
    		& \ \ \ \ \ \frac{d}{dt} \left( \begin{aligned}
                a \\ b
                            \end{aligned}\right)
      = -iH^{'} \left( \begin{aligned}
                a \\ b
            \end{aligned}\right).
        \end{aligned}   
    \end{equation}
    with a dynamic matrix expressed as
    \begin{equation}
    \label{AppEq2}
        H^{'}=\left(
			\begin{aligned}
				\omega_{1}+iG-i\gamma|a|^{2} \ \ \ \ \ \ \ \ J \ \ \\
				J \ \ \ \ \ \ \ \ \ \ \omega_{2}-i\kappa
			\end{aligned}\right).
    \end{equation}
    Here, $H^{'}$ captures the nonlinear gain-loss dynamics of the system, identical to Eq. \ref{Eq3}. The eigenfrequencies of this system, which depend on the amplitude, are given by
    \begin{equation}
        {{\omega }_{\pm }}=\omega_{r}-i\frac{{G}-\gamma|a|^{2}-\kappa}{2}\pm \frac{1}{2}\sqrt{4{J^{2}}+{{(\Delta-i({G}-\gamma|a|^{2}+\kappa))}^{2}}},
    \end{equation}
    where $\omega_{r}=({\omega }_{1}+\omega_{2})/2$ and $\Delta=\omega_{2}-\omega_{1}$.

    For the system to achieve a steady amplitude, the eigenfrequencies must be purely real, enforcing the condition $\textbf{Im}[\omega(|a_{c}|)_{\pm}] = 0$\cite{Zhang}. This requirement ensures that the hybridized modes at steady state exhibit neither net gain nor net loss. For the case, $\omega_{1}\neq\omega_{2}$, the two eigenfrequencies always correspond to different amplitudes, and the system stabilizes at the amplitude where one eigenmode remains real while the other becomes dissipative. This self-selection mechanism, in which the system naturally selects the stable mode, was first proposed by Yao et al. \cite{GDP}.

    Once the coupled system relaxes to its final steady state, where $G>\kappa$, the system will reach a steady amplitude under the condition $\textbf{Im}[\omega(|a_{c}|)_{\pm}] = 0$, as discussed previously. This leads to the following relationship,
    \begin{equation}
        \beta=G-\gamma|a|^{2}=\kappa,
        \label{AppEq5}
    \end{equation}
    which indicates that the net gain $\beta$ of the gain-driven mode is exactly balanced by the loss in the lossy mode. By substituting Eq. \ref{AppEq5} into Eq. \ref{AppEq2}, we can derive a linear gain-loss coupled system, described by the following matrix
    \begin{equation}
		\begin{aligned}
			H=H^{'}(|a|)=&\left(
			\begin{aligned}
				\omega_{0}+i\beta \ \ \ \ \ \ J \ \ \ \ \\
				 \ \ \ \ J \ \ \ \ \ \ \ \omega_{0}-i\beta
			\end{aligned}\right),
		\end{aligned}
    \end{equation}
    which matches the linear gain-loss coupled Hamiltonian as presented in Eq. \ref{Eq2}. This transformation is derived under nonlinearity, indicating that the nonlinear gain-loss coupled system can also be PT-symmetric.

\subsection{Linear dynamics}
    In the linear regime, the system is typically considered under the condition $\omega_{1}=\omega_{2}=\omega_{0}$. Here, the van der Pol nonlinearity can be neglected for the small amplitude oscillations. When the system operates in the dissipative regime, where the loss exceeds the gain, $\kappa>G$, the oscillations do not grow to high amplitudes, allowing the van der Pol term to be neglected, $\gamma=0$. The dynamic matrix of the system then simplifies to
    \begin{equation} \label{AppEq3}
        H^{'}=\left(
			\begin{aligned}
				\omega_{0}+iG \ \ \ \ \ \ J \ \ \ \ \\
				\ \ \ \ J \ \ \ \ \ \ \ \omega_{0}-i\kappa
			\end{aligned} \right).
    \end{equation}
    To uncover the system’s hidden PT-symmetry, we define a reference parameter $\chi=(G-\kappa)/2$. Using this reference, Eq. \ref{AppEq3} can be rewritten as\cite{ozdemir2019parity} 
    \begin{equation}
		\begin{aligned}
			H^{'}=H+i\chi\mathbb{I}
		\end{aligned}
    \end{equation}
    where the matrix $H$ is given by
    \begin{equation}
		\begin{aligned}
			H=\left(
			\begin{aligned}
				\omega_{0}+i\beta \ \ \ \ \ \ J \ \ \ \ \\
				 \ \ \ \ J \ \ \ \ \ \ \ \omega_{0}-i\beta
			\end{aligned}\right),
		\end{aligned}
        \label{AppEq4}
    \end{equation}
    and $\beta=(G+\kappa)/2$ represents the effective gain. Here, $\mathbb{I}$ is the identity matrix. It can be shown that with the reference $i\chi$, the matrix $H$ exhibits PT-symmetry, as follows:
    \begin{equation}
        H  \overset{{\scriptscriptstyle \textbf{P}}}{\Longrightarrow}
        \left(\begin{aligned}
				\tilde{\omega}_{l} \ \ \ \ \ \ J \ \ \ \ \\
				 \ \ \ \ J \ \ \ \ \ \ \ \tilde{\omega}_{g}
			\end{aligned}\right) \overset{{\scriptscriptstyle \textbf{T}}}{\Longrightarrow}
        \left(\begin{aligned}
				\tilde{\omega}_{l}^{*} \ \ \ \ \ \ J \ \ \ \ \\
				 \ \ \ \ J \ \ \ \ \ \ \ \tilde{\omega}_{g}^{*}
			\end{aligned}\right)=H.
    \end{equation}
    where the parity transformation ($\textbf{P}$) exchanges the two modes, and the time-reversal transformation ($\textbf{T}$) involves complex conjugation. Together, these transformations preserve the structure of the dynamic matrix, demonstrating its PT-symmetry.

    When the gain exceeds the loss, $G>\kappa$, the system can still be described by Eq. \ref{AppEq3}, but this now corresponds to the initial transient state. In this regime, the oscillations are activated by the gain but are far from reaching their steady amplitude. Although the final steady state governs the emission power of the system, the initial transient state plays a critical role in determining whether the system will be activated by gain. By applying the appropriate reference, its dynamic matrix can be further gauged into the form of Eq. \ref{AppEq4}.
    
\section{Hamiltonian of virtual gain-loss coupled system}
    As discussed in Sec.\ref{sec2C}, complex decaying signals can induce a virtual gain effect in the system without requiring any active components. Here, we examine the dynamics of two coupled resonators, characterized by coupling strength $J$ and intrinsic losses ${{\kappa }_{L1}}$ and ${{\kappa }_{L2}}$. The system's Hamiltonian is given by $\hat{H}={{\omega }_{01}}{{a}^{\dagger }}a+{{\omega }_{02}}{{b}^{\dagger }}b+J \left( {{a}^{\dagger }}b+a{{b}^{\dagger }} \right)-i{{\kappa }_{L1}}{{a}^{\dagger }}a-i{{\kappa }_{L2}}{{b}^{\dagger }}b$, where ${{a}^{\dagger }}$ and $a$ (${{b}^{\dagger }}$ and $b$) are the creation and annihilation operators for the first (second) resonator mode. In the linear regime, where quantum fluctuations can be neglected, this Hamiltonian can be reduced to a $2\times 2$ matrix using coupled mode theory. Under these conditions, the system's behavior is described by the classical field amplitudes in each resonator rather than quantum operators. Focusing on the expectation values $\langle a\rangle $ and $\langle b\rangle $ for the two resonator modes, the Hamiltonian reduces to a non-Hermitian matrix $\hat{H}=\left( \begin{matrix}
       {{\omega }_{01}}-i{{\kappa }_{L1}} & J   \\
       J  & {{\omega }_{02}}-i{{\kappa }_{L2}}  \\
    \end{matrix} \right)$. The corresponding temporal coupled-mode theory equations are
    \begin{equation}
        \mathbf{\dot{\hat{a}}}=-i\left( \begin{matrix}
       {{\omega }_{01}}-i({{\kappa }_{L1}}+{{\kappa }_{E1}}) & J   \\
       J  & {{\omega }_{02}}-i({{\kappa }_{L2}}+{{\kappa }_{E2}})  \\
    \end{matrix} \right)\mathbf{\hat{a}}+\mathbf{\hat{s}},
    \end{equation}
    where $\mathbf{\hat{a}}\equiv \left({{a}_{1}}, {{a}_{2}}\right)^{T}$ is the amplitude vector and $\mathbf{\hat{s}}\equiv \left(\sqrt{2{{\kappa }_{E1}}}{{s}_{1}}, \sqrt{2{{\kappa }_{E2}}}{{s}_{2}} \right)^{T}$ is the excitation vector.

    For the complex frequency excitation ${{s}_{1,2}}(t)={{s}_{0}}{{e}^{-i({{\omega }_{r}}-i{G_{1,2}})t}}={{s}_{0}}{{e}^{-i{{\omega }_{r}}t}}{{e}^{{-G_{1,2}}t}}$, a process similar to that described in Appendix \ref{APPvirtualG} leads to the effective inclusion of the complex frequency in the system's dynamics. This modifies the Hamiltonian to 
    \begin{equation}
        {{\hat{H}}_{m}}=\left( \begin{matrix}
       {{\omega }_{01}}-i({{\kappa }_{L1}}-{G_{1}}) & J   \\
       J  & {{\omega }_{02}}-i({{\kappa }_{L2}}-G_{2})  \\
    \end{matrix} \right).
    \end{equation} 
    Assuming the intrinsic eigenfrequencies of both modes are equal, ${{\omega }_{01}}={{\omega }_{02}}={{\omega }_{0}}$, and simplifying by considering the first mode to be lossless with complex excitation only from this side, the eigenvalue calculation of the Hamiltonian yields, 
    \begin{equation}
        {{\omega }_{\pm }}={{\omega }_{0}}-i\frac{{G}-{{\kappa }_{2}}}{2}\pm \frac{1}{2}\sqrt{4{J^{2}}-{{({G}+{{\kappa }_{2}})}^{2}}}.
    \end{equation}
    In the specific case where $G={{\kappa }_{2}}$, the system's eigenvalues become
    \begin{equation}
        {{\omega }_{VPT}}={{\omega }_{0}}\pm \sqrt{{J^{2}}-G^{2}}
    \end{equation}
    If the coupling strength exceeds the loss ($J >{\kappa_{2}}$), the eigenvalues are purely real, indicating the presence of a PT-symmetric phase. In this scenario, the system remains purely passive, and PT symmetry is achieved through complex frequency excitation, giving rise to what is known as the virtual PT symmetry state\cite{VPTS2020, Alex2020, Alex2021}. When the coupling is fine-tuned to $J ={{\kappa }_{2}}=G$, the eigenvalues and eigenvectors coalesce, leading to the EP state.

%

\end{document}